\documentclass[journal]{IEEEtran}

\usepackage[english]{babel}
\usepackage{cite}
\usepackage{graphicx}
\usepackage[lowtilde]{url}
\usepackage[caption=false,font=footnotesize]{subfig}
\usepackage{amssymb}
\usepackage{amsmath}
\usepackage{multirow}
\usepackage[euler]{textgreek}
\usepackage{booktabs}
\usepackage{array}
\usepackage[ruled,norelsize]{algorithm2e}
\usepackage{ragged2e}
\usepackage{mathtools}
\usepackage{xcolor}

\makeatletter
\long\def\@makecaption#1#2{\ifx\@captype\@IEEEtablestring%
\footnotesize\begin{center}{\normalfont\footnotesize #1}\\
{\normalfont\footnotesize\scshape #2}\end{center}%
\@IEEEtablecaptionsepspace
\else
\@IEEEfigurecaptionsepspace
\setbox\@tempboxa\hbox{\normalfont\footnotesize {#1.}~~ #2}%
\ifdim \wd\@tempboxa >\hsize%
\setbox\@tempboxa\hbox{\normalfont\footnotesize {#1.}~~ }%
\parbox[t]{\hsize}{\normalfont\footnotesize \noindent\unhbox\@tempboxa#2}%
\else
\hbox to\hsize{\normalfont\footnotesize\hfil\box\@tempboxa\hfil}\fi\fi}
\makeatother

\newcommand{\RN}[1]{%
  \textup{\uppercase\expandafter{\romannumeral#1}}%
}

\makeatletter
\newcommand{\removelatexerror}{\let\@latex@error\@gobble}
\makeatother

\newcolumntype{C}[1]{>{\centering\let\newline\\\arraybackslash\hspace{0pt}}m{#1}}

\begin{document}
\title{Assessing the Quality-of-Experience of Adaptive Bitrate Video Streaming}
\author{Zhengfang~Duanmu,~\IEEEmembership{Student Member,~IEEE,}
        Wentao~Liu,~\IEEEmembership{Student Member,~IEEE,}\\
        Zhuoran~Li,~\IEEEmembership{Student Member,~IEEE,}
        Diqi~Chen,\\
        Zhou~Wang,~\IEEEmembership{Fellow,~IEEE,}
        Yizhou~Wang,
        and~Wen~Gao,~\IEEEmembership{Fellow,~IEEE}
\thanks{Z. Duanmu, W. Liu, Z. Li, and Z. Wang are with the Department of Electrical and Computer Engineering, University of Waterloo, Waterloo, ON N2L 3G1, Canada (e-mail: \{zduanmu, w238liu, z777li, zhou.wang\}@uwaterloo.ca).}
\thanks{D. Chen is with Institute of Computing Technology, Chinese Academy of Sciences, Beijing, 100190, China (e-mail: cdq@pku.edu.cn).}
\thanks{Y. Wang, and W. Gao are with the School
of Electronic Engineering and Computer Science, Peking University, Beijing, 100871, China (e-mail: \{yizhou.wang, wgao\}@pku.edu.cn).}
}


\maketitle

\begin{abstract}
The diversity of video delivery pipeline poses a grand challenge to the evaluation of adaptive bitrate (ABR) streaming algorithms and objective quality-of-experience (QoE) models.
Here we introduce so-far the largest subject-rated database of its kind, namely WaterlooSQoE-IV, consisting of 1350 adaptive streaming videos created from diverse source contents, video encoders, network traces, ABR algorithms, and viewing devices.
We collect human opinions for each video with a series of carefully designed subjective experiments.
Subsequent data analysis and testing/comparison of ABR algorithms and QoE models using the database lead to a series of novel observations and interesting findings, in terms of the effectiveness of subjective experiment methodologies, the interactions between user experience and source content, viewing device and encoder type, the heterogeneities in the bias and preference of user experiences, the behaviors of ABR algorithms, and the performance of objective QoE models.
Most importantly, our results suggest that a better objective QoE model, or a better understanding of human perceptual experience and behaviour, is the most dominating factor in improving the performance of ABR algorithms, as opposed to advanced optimization frameworks, machine learning strategies or bandwidth predictors, where a majority of ABR research has been focused on in the past decade.
On the other hand, our performance evaluation of 11 QoE models shows only a moderate correlation between state-of-the-art QoE models and subjective ratings, implying rooms for improvement in both QoE modeling and ABR algorithms.
The database is made publicly available at: \url{https://ece.uwaterloo.ca/~zduanmu/waterloosqoe4/}.
\end{abstract}


\IEEEpeerreviewmaketitle

\section{Introduction}
Since the ratification of Dynamic Adaptive Streaming over HTTP (DASH) standard in 2011~\cite{stockhammer2011dynamic}, video distribution service providers have invested significant effort in the transition from the conventional connection-oriented video transport protocols towards DASH due to its ability to traverse network address translations and firewalls, reliability to deliver video packets, flexibility to react to volatile network conditions, and efficiency in reducing the server workload.
DASH video players aim to optimize viewers' quality-of-experience (QoE) subject to bandwidth constraints by adaptively selecting download bitrate from a pre-defined set of media streams.
Adaptive bitrate (ABR) streaming algorithms, that determine the bitrate of the next media segment, are deliberately left open for optimization.

The past decade has seen a rapid advancement of ABR algorithms~\cite{li2014probe,huang2015buffer,mok2012qdash,sun2016cs2p,spiteri2016bola,jiang2014improving,yin2015control,mao2017neural,chiariotti2016online,claeys2013design,van2015learning,akhtar2018oboe}, from na\"ive linear bandwidth prediction and greedy bitrate selection to sophisticated data-driven throughput estimation and long-term QoE optimization.
With a variety of ABR logics available, how to fairly evaluate their performance becomes pivotal. 
The conventional approach measures the performance of existing ABR logics with objective metrics such as average bitrate, rebuffering time, join time, bitrate switches in well-characterized communication pipelines.
Regardless of the inevitably more complex real-world communication environment, the reliability of these isolated QoE measures is often controversial~\cite{duanmu2017qoe}.
As the human visual system is the ultimate receiver of adaptive streaming videos, subjective evaluation is the most straightforward and reliable approach to evaluate the ABR techniques.
A comprehensive subjective user study not only helps better understand human perceptual QoE, but also creates the basis to evaluate, compare, and optimize ABR algorithms and objective QoE models.

\begin{table*}[t]
\centering
\caption{Comparison of Publicly Available QoE Databases for HTTP-based Adaptive Video Streaming}
\label{tab:databases}
\begin{tabular}{c|c c c c c c}
\toprule
	\multirow{2}{*}{Database} & \# of Source & \# of & \# of Network & \# of ABR & Viewing & \# of Test \\
	 		 & Videos & Encoders & Traces & Algorithms & Device & Videos \\\hline
    LIVEMVQA & 10 & 1 & 0 & - & Phone, Tablet & 300 \\
    LIVEQHVS & 3 & 1 & 0 & - & HDTV & 15 \\
    LIVEMSV & 24 & 1 & 0 & - & Phone & 176 \\
    LIVE-NFLX-I & 14 & 1 & 0 & - & Phone & 112 \\
    LIVE-NFLX-II & 15 & 1 & 7 & 4 & HDTV & 420 \\ 
    WaterlooSQoE-I & 20 & 1 & 0 & - & HDTV & 200 \\
    WaterlooSQoE-II & 12 & 1 & 0 & - & HDTV & 588 \\
    WaterlooSQoE-III & 20 & 1 & 13 & 6 & HDTV & 450 \\\hline
    WaterlooSQoE-IV & 5 & 2 & 9 & 5 & Phone, HDTV, UHDTV & 1350 \\
\bottomrule
\end{tabular}
\end{table*}

Nevertheless, a large-scale subjective quality assessment experiment faces two primary practical challenges.
First, given the high dimensionality of streaming videos, it is prohibitively difficult to perform an exhaustive subjective evaluation on all possible visual stimuli.
As a result, a faithful evaluation of ABR algorithms should cover well-controlled, diverse, representative, and realistic streaming videos, which cannot be realized by handcrafted test samples or a small number of random samples of real-world streaming videos in the wild.
Second, traditional subjective experiment methodologies that are designed for a small number of short video clips may not be appropriate in the evaluation of large-scale prolonged streaming videos.
Given the extended period of experiment, subjects may not attend to the whole visual stimuli due to the loss of interest and the limited mental capacity.
How to keep viewers focused on the test stimuli without intrusively influencing the reliability of QoE ratings remains an open question.
We tackle these problems by carefully walking through the selections of each of the key components in the ABR streaming process, from source contents, encoding profiles, network traces, ABR algorithms, viewing devices, testing environment setups, to subjective testing methodologies.
Our work leads to so far the most comprehensive streaming video QoE database, which gives us an unique opportunity to move on further and explore many unresolved problems in streaming video QoE and ABR streaming algorithms, resulting in a series of new findings and observations.

The main contributions of our work include:
\begin{itemize}
    \item A large-scale video database, which we name Waterloo Streaming QoE database IV (WaterlooSQoE-IV), consisting of 1350 realistic streaming videos generated from a variety of transmitters, channels, and receivers.
    The dataset covers a wide spectrum of ABR techniques.
    To enable fair comparison, we optimize each ABR algorithm over an independent set of training videos.
    \item A large subjective experiment conducted in a well-controlled laboratory environment.
    We carefully design our experimental protocols to collect the mean opinion score (MOS) on three viewing devices and verify its reliability.
    Additional experiment is performed to align MOSs obtained from different viewing sessions.
    \item An in-depth analysis of the influencing factors.
    We find interesting relationship between the viewing condition and perception of rebuffering/quality adaptation, which has not been observed in previous studies.
    We also identify two types of user heterogeneity in QoE perception.
    \item An extensive evaluation of ABR algorithms.
    We show with surprise that state-of-the-art algorithms using advanced optimization schemes may not outperform the na\"ive rate-based algorithm, while significant gain can be obtained with a perceptual motivated objective function.
    \item A comprehensive evaluation of objective QoE models.
    We calibrate $11$ QoE models on an independent set of training data to assess their performance in terms of the correlation with human opinions.
    We make the implementations of the models publicly available.
\end{itemize}

\section{Related Work}
\subsection{Comparative Study of ABR Algorithms}
The past decade has witnessed a plethora of comparative studies of ABR algorithms, but how to validate and compare them has been a major challenge.
Existing evaluation methods can be categorized as objective evaluation, indirect subjective evaluation, and direct subjective evaluation.
Most works employ average bitrate, rebuffering time, join time, and bitrate switches as proxies of subjective QoE~\cite{nam2014youslow,yin2015control,mao2017neural,wang2019resa,huang2019hindsight,Yan2020puffer}.
However, these objective metrics neglect the heterogeneity in source contents, video codecs, viewing devices, and individual preferences.

Very few ABR developers validate the algorithms with real-world deployment, where user engagement (\textit{e.g.} percentage of viewing video) is considered as an indirect but meaningful measure of QoE~\cite{dobrian2011understanding,krishnan2013video}.
Several obstacles prevent the broad utility of this approach.
In addition to the proprietary nature of ABR infrastructures, the evaluation procedure is often executed without variable control, making the results inevitably unconvincing.
Most importantly, the equivalence between user engagement and QoE is perplexed by user interests, random exit, and unattended video playback.

Our work is most relevant to three previous studies.
Bampis~\textit{et al.}~\cite{bampis2018towards} and our previous studies~\cite{rodriguez2014impact,duanmu2018quality} collect human opinion scores of realistic streaming videos generated by a set of ABR algorithms.
Aside from the fact that the ABR algorithms under evaluation are severely dated (proposed before 2015), the scope of these studies are limited to three dimensions including encoding, network throughput and the choice of ABR algorithms.

\begin{figure*}[t]
    \centering
    \captionsetup{justification=centering}
    \subfloat[Slides]{\includegraphics[width=0.2\textwidth]{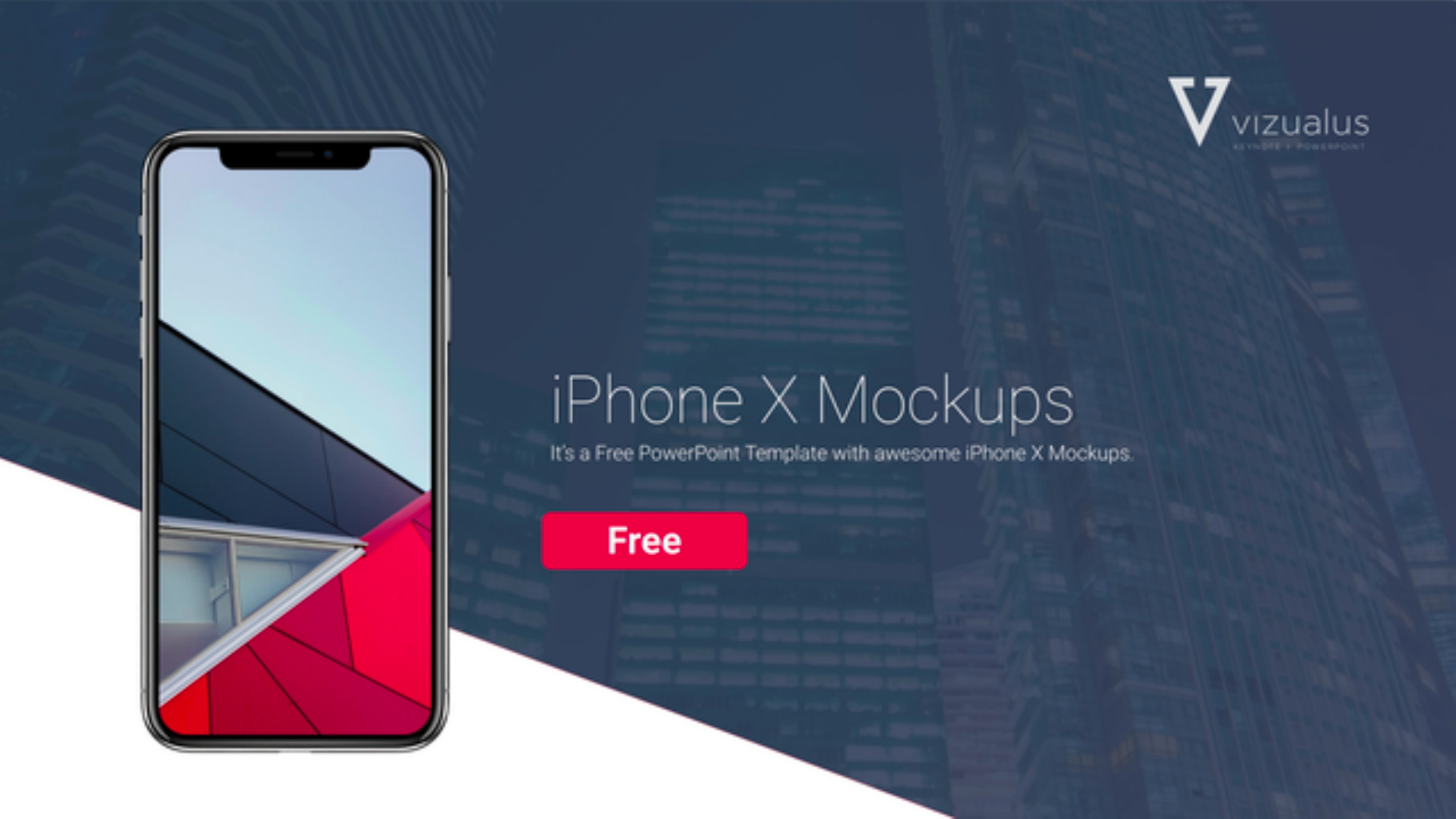}}
    \subfloat[Game]{\includegraphics[width=0.2\textwidth]{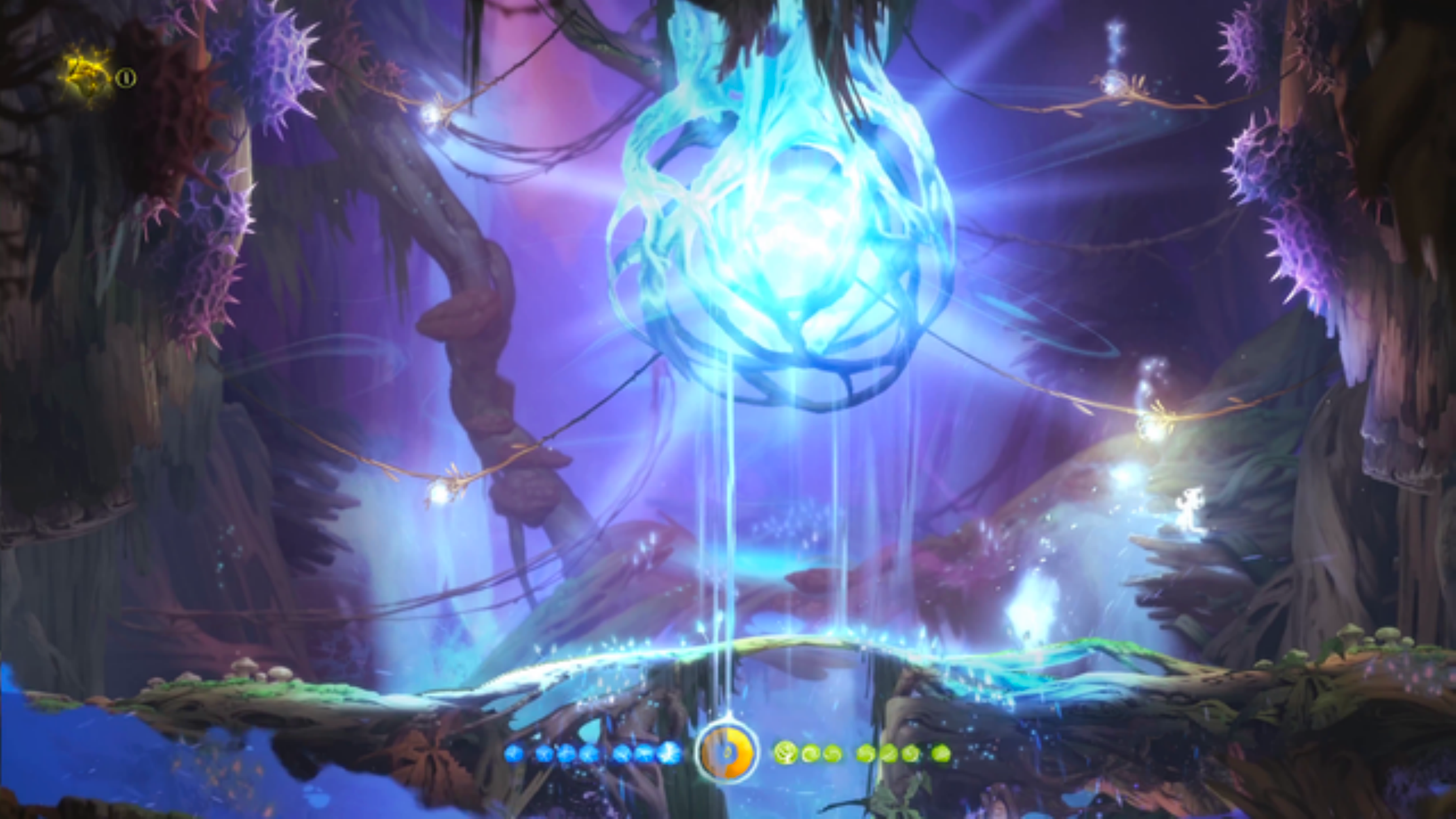}}
    \subfloat[Movie]{\includegraphics[width=0.2\textwidth]{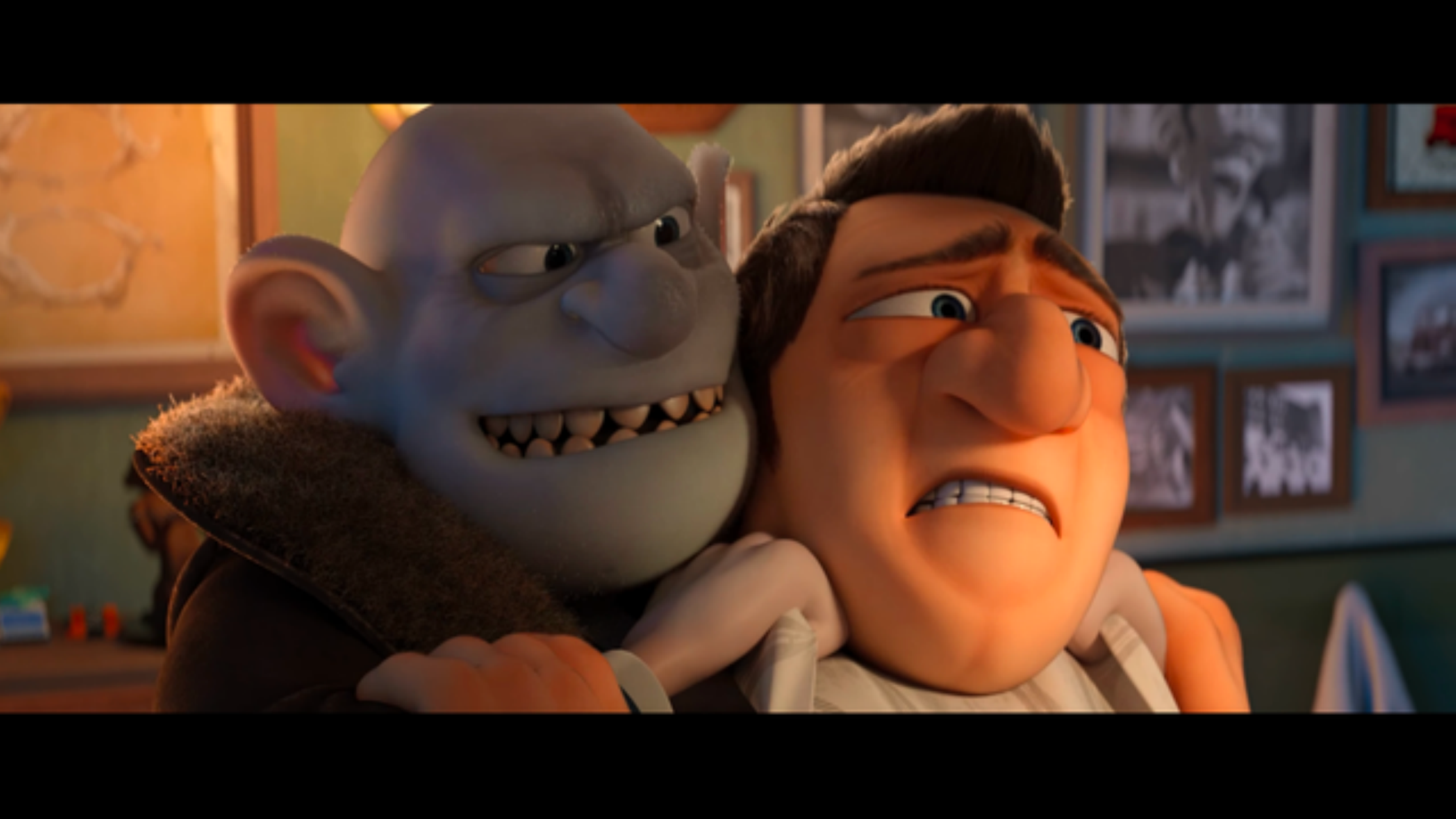}}
    \subfloat[Nature]{\includegraphics[width=0.2\textwidth]{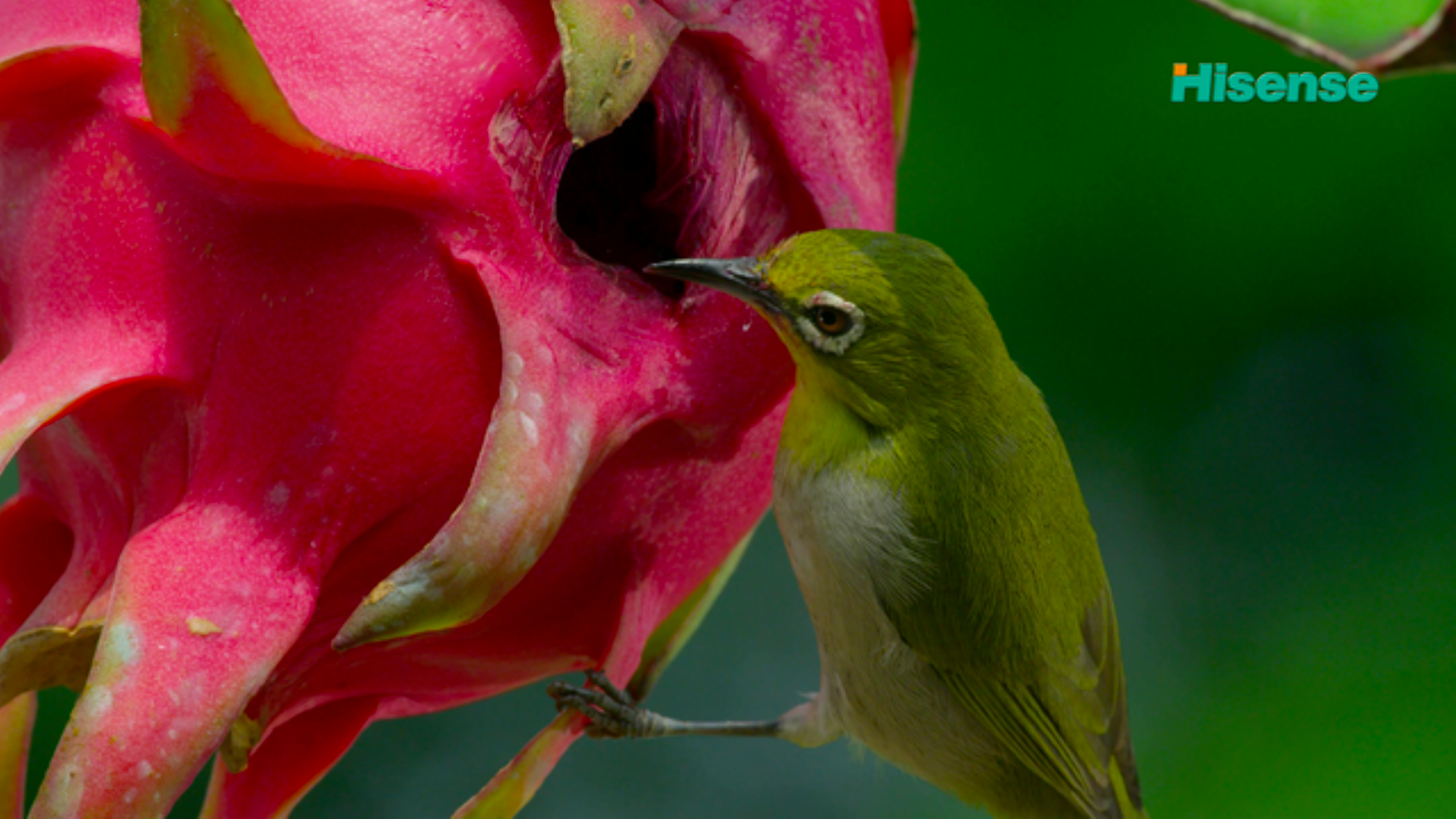}}
    \subfloat[Sport]{\includegraphics[width=0.2\textwidth]{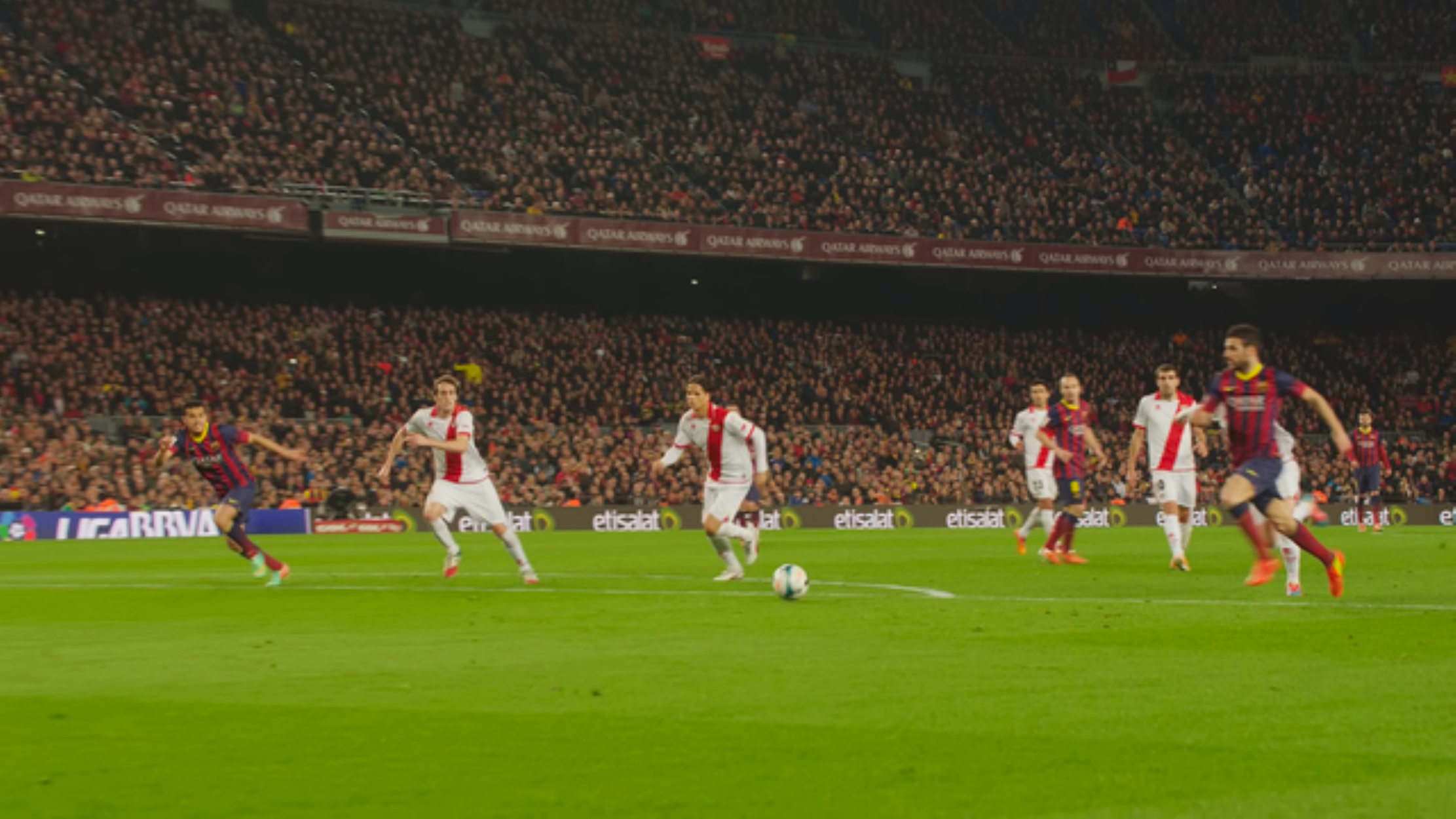}}
    \caption{Snapshots of video sequences.}\label{fig:testseqs}
\end{figure*}

\subsection{Streaming Video QoE Datasets}
A number of subject-rated QoE datasets have served as training and evaluation benchmarks for objective QoE models.
In 2012, Moorthy \textit{et al.}~\cite{moorthy2012video} presented one of the first attempts to analyze the subjective QoE response to adaptive streaming videos.
The constructed LIVE mobile video quality assessment database (LIVEMVQA) consists of 200 short videos evaluated by over 30 human subjects on a smart phone, as well as 100 distorted videos evaluated by 17 subjects on a tablet.
The test videos are contaminated by isolated distortions including H.264 compression, rebuffering, frame drop, quality adaptation, and wireless channel packet-loss.

The LIVE Mobile Stall Video database (LIVEMSV)~\cite{ghadiyaram2014study} focuses on the rebuffering experience, which has been widely accepted as the most annoying distortion in adaptive streaming.
The dataset contains a total of 176 videos generated from 24 reference videos with 26 hand-crafted stalling events.
The quality labels are obtained from 54 subjects who viewed the videos on Apple iPhone devices.

The Waterloo Streaming QoE database-I (WaterlooSQoE-I)~\cite{duanmu2016sqi} is another rebuffering-centric dataset covering diverse video contents.
Each reference video is encoded into three bitrates with H.264 encoder and then a rebuffering event is simulated at either the beginning or the middle point of the encoded sequences.
In total, 200 test videos are generated and displayed on high definition television (HDTV) for subjective evaluation.

The LIVE QoE database for HTTP-based Video Streaming (LIVEQHVS)~\cite{chen2014modeling} contains three 300-second long temporally incoherent reference videos concatenated from eight short video clips.
For each reference video, 5 bitrate-varying videos are constructed by adjusting the encoding bitrate of H.264 video encoder, resulting in a relatively small set of 15 quality-varying videos.
A subjective study is conducted to measure the QoE of test stimuli on a 58 inch Panasonic HDTV plasma monitor.

The Waterloo Streaming QoE database II (WaterlooSQoE-II)~\cite{duanmu2018ect} is a dataset of 588 video clips with variations in compression level, spatial resolution, and frame-rate, featured in the segment level QoE evaluation.
All videos are displayed at their actual pixel resolution on an HDTV for subjective QoE rating.
The inclusion of segment level quality ratings and the broad range of adaptation patterns make it ideal for the calibration of objective quality switching experience models.

A common issue with these streaming video QoE datasets is the isolated distortion types such as rebuffering and quality adaptation.
The LIVE-NFLX-I database~\cite{bampis2017qoe} is developed to understand the influence of mixtures of streaming video distortions in subjective QoE.
The database involves 112 distorted videos with 8 handcrafted playout patterns evaluated by over 55 human subjects on a mobile device.
Unfortunately, only a fraction of videos are made publicly available for license issues.

In spite of the authors' effort in designing meaningful distortion patterns, the test sequence in the aforementioned datasets are handcrafted, significantly deviating from real-world streaming video distributions.
Motivated by the limitation, the LIVE-NFLX-II~\cite{bampis2018towards} and the Waterloo Streaming QoE database-III (WaterlooSQoE-III)~\cite{duanmu2018quality} provide 420 and 450 realistic streaming videos, respectively.
These two datasets employ network emulator and realistic ABR algorithms for video delivery, based on which the streaming videos are recorded and reconstructed.
Both experiments collect subjective QoE scores on a 1080p computer monitor.
A summary of the aforementioned databases are given in Table~\ref{tab:databases}.

All of the above studies suffer from the following problems: (1) the datasets are limited in size; (2) the prevalent 4K resolution is missing; (3) advanced video encoders are not included; and (4) the human studies are conducted on few out-dated devices.
The above issue motivates us to build a new database for QoE research, which aims to provide more diverse and realistic streaming videos spanning a broad range of transmitters, channels, and receivers.

\section{Adaptive Streaming Video Database Construction}
In this section, we first describe the construction of the proposed WaterlooSQoE-IV database including the source material collection and the simulation experiment setup.
We then present the details of the subjective experiment for collecting human annotations.

\begin{table}[t]
\centering
\caption{Information of Reference Videos}
\label{tab:testseqs}
\scalebox{0.95}{
\begin{tabular}{c|c|c}
\toprule
    Name & Frame Rate & Description\\\hline
    Slides & 30 & screen content, static, scene switch \\
    Game & 30 & animation, high motion \\
    Movie & 24 & computer generated, high motion, coherent scene \\
    Nature & 30 & natural, animal, scene switch \\
    Sport & 30 & human, high motion, view change \\
\bottomrule
\end{tabular}
}
\end{table}

\subsection{Database Construction}\label{sec:lssvd}
\noindent\textbf{Source videos:} We select five high-quality 4K creative commons licensed videos from the Internet, which span a diverse set of content genres, including screen content, video game, movie, natural scene and sport.
There are different types of camera motion, including static (\textit{e.g.} Slides, Game and Nature) and complex scenes taken with a moving camera, with panning and zooming (\textit{e.g.} Movie and Sport).
To make sure that the videos are of pristine quality, we carefully inspect each of the videos multiple times by zooming in and remove those videos with visible distortions.
The detailed specifications of those videos are listed in Table~\ref{tab:testseqs} and a screenshot from each video is included in Figure~\ref{fig:testseqs}.
To accommodate the limited subjective experiment capacity, we cut a 32-second video clip from each source content.
It should be noted that the duration of test sequences is longer than the settings used in the existing streaming video QoE datasets.
This choice is in accordance with many recent studies~\cite{chen2014modeling,bampis2018towards}, which suggest that longer videos of up to 30 seconds may be required to be able to test the impact of switching patterns.

\noindent\textbf{Encoding profiles:} Using the aforementioned sequences as the source, each video is encoded with H.264~\cite{wiegand2003h264} and HEVC~\cite{sullivan2012hevc} encoders into 13 representations using the bitrate ladder shown in Table~\ref{tab:reps} to cover different quality levels.
The choices of bitrate levels are based on Netflix's recommendation~\cite{netflix2015bitrateladder} while the last two representations are appended to the original bitrate ladder to cover the high-quality representations suggested in Apple's recommendation~\cite{apple2016bitrateladder}.
Despite the recent development in content adaptive bitrate ladder generation~\cite{toni2015optimal,de2016complexity,duanmu2020ramct}, there has been no widely accepted per-title encoding strategy.
Furthermore, some ABR algorithms only accept a fixed set of encoding profiles as input~\cite{mao2017neural}.
We segment the test sequences with GPAC's MP4Box~\cite{le2007gpac} with a segment length of $4$ seconds.
Since some testing ABR algorithms rely on chunk-level bitrate and presentation quality scores in the bitrate selection, we pre-compute and embed them as the attributes of SegmentURL~\cite{iso2012dash} in the manifest file that describes the specifications of the video.

\begin{figure}[t]
\centering
\includegraphics[width=0.8\linewidth]{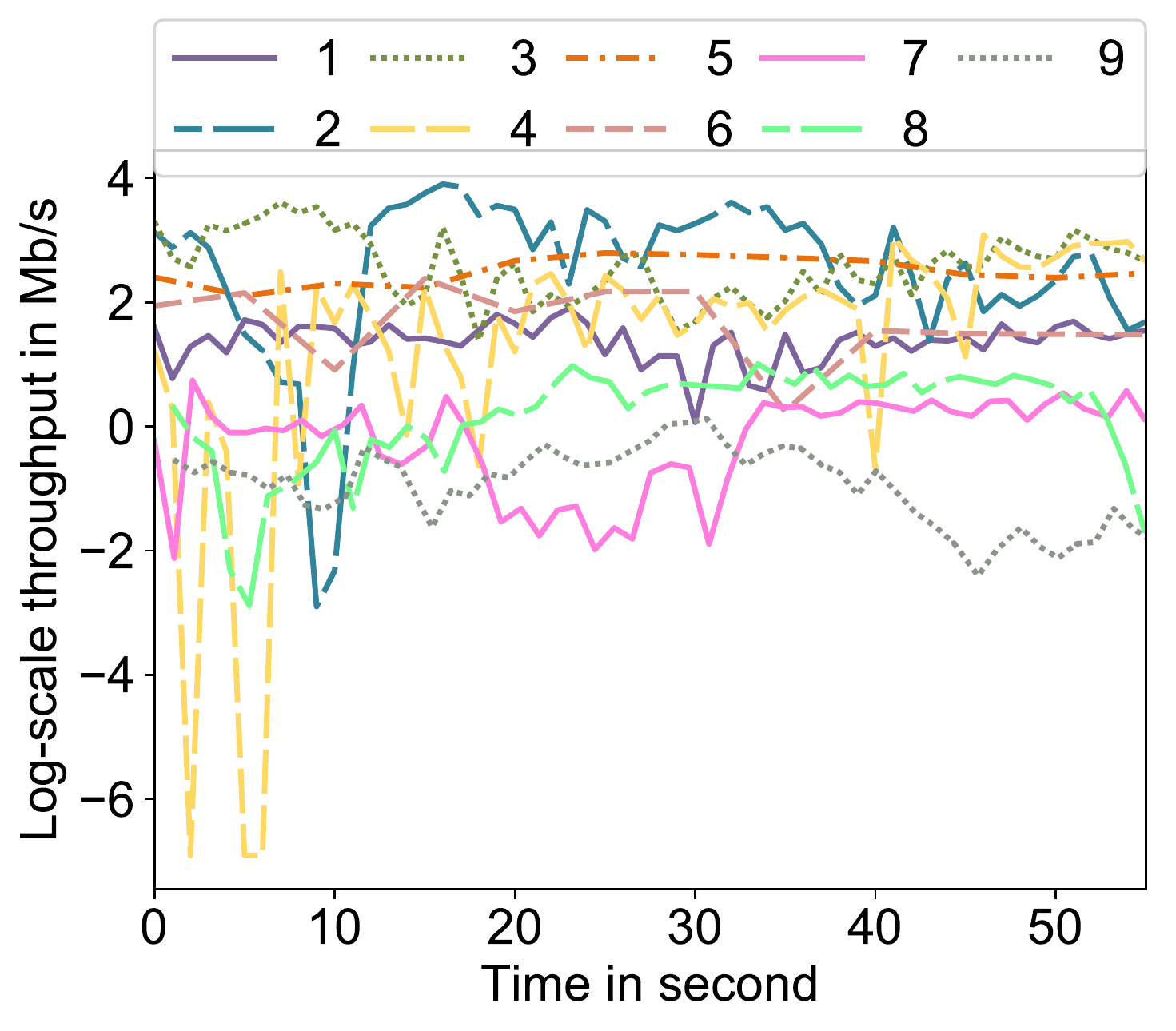}
\caption{Network traces used in the subjective experiment.}
\label{fig:network_trace_plot}
\end{figure}

\noindent\textbf{Network traces:} To evaluate ABR algorithms on realistic network conditions, we employ the combination of several existing datasets: a broadband dataset (FCC)~\cite{fcc2016dataset}, a 3G mobile dataset (HSDPA)~\cite{riiser2013hdspa}, and a 4G mobile dataset (Belgium)~\cite{vander2016belgium}.
The FCC dataset contains more than 1 million throughput traces, each of which records the average throughput over 2100 seconds at a granularity of 5 seconds.
We select traces by randomly cutting from the ``web get'' category in the August 2016 collection, each with a duration of 55 seconds.
The HSDPA dataset comprises 30 minutes of throughput measurements, collected from mobile devices that were streaming video while in transit.
The Belgium dataset consists of 40 LTE bandwidth traces recorded along several routes in and around the city of Ghent at a 1-second granularity.
To match the duration of our selected FCC traces, we generate traces using a sliding window across the HSDPA dataset and the Belgium dataset respectively.
To avoid scenarios where bitrate selection is trivial, \textit{i.e.}, situations where picking the minimum bitrate still causes serious stalling events, we only considered the original traces whose average throughput is greater than $0.2$ Mb/s.
We pick nine network traces from the collected corpus.
As shown in Figure~\ref{fig:network_trace_plot}, the selected traces are approximately 55 seconds in duration and have varying network behaviors.
Some network traces are likely to cause sudden bitrate/quality changes and rebufferings even if the average bandwidth is relatively high.

\begin{table}[t]
\centering
\caption{Encoding Ladder of Video Sequences}\label{tab:reps}
\scalebox{0.8}{
    \begin{tabular}{c|c c|c|c c}
    \toprule
        Index & Resolution & Bitrate & Index & Resolution & Bitrate\\\hline
        1 & 320$\times$180 & 235 Kb/s & 8 & 1280$\times$720 & 3000 Kb/s \\
        2 & 384$\times$216 & 375 Kb/s & 9 & 1920$\times$1080 & 4300 Kb/s\\
        3 & 512$\times$288 & 560 Kb/s & 10 & 1920$\times$1080 & 5800 Kb/s\\
        4 & 512$\times$288 & 750 Kb/s & 11 & 2560$\times$1440 & 8100 Kb/s\\
        5 & 640$\times$360 & 1050 Kb/s & 12 & 3840$\times$2160 & 11600 Kb/s\\
        6 & 960$\times$540 & 1750 Kb/s &  13 & 3840$\times$2160 & 16800 Kb/s\\
        7 & 1280$\times$720 & 2350 Kb/s & & &\\
    \bottomrule
    \end{tabular}
}
\end{table}

\noindent\textbf{ABR algorithms:} Exhaustive comparison of all the existing ABR algorithms is difficult as it involves optimizing over an infinite-dimensional functional space. To this end, we evaluate the following ABR algorithms, ranging from the de facto rate-based algorithm to the state-of-the-art algorithms:
\begin{itemize}
    \item Rate-Based (RB): RB~\cite{iso2012dash} picks the maximum available bitrate below the throughput predicted by the arithmetic mean of past 5 chunks.
    \item Buffer-Based (BB): We employed the function suggested by Huang \textit{et al.}~\cite{huang2015buffer}, where bitrate is chosen as a piece-wise linear function of buffer occupancy. The lower reservoir and cushion are set to 5 and 10 seconds, respectively.
    \item FastMPC: FastMPC~\cite{yin2015control} uses both buffer occupancy observations and throughput predictions using harmonic mean of the past 5 chunks to select bitrate which maximizes a bitrate-based QoE metric over a horizon of 5 future chunks. The optimization problem is solved offline and its solution is stored as a lookup table. We use 100 bins for throughput prediction, 100 bins for buffer level, and 13 bins for the past bitrate level.
    \item Pensieve: Pensieve~\cite{mao2017neural} is a reinforcement learning-based ABR algorithm.
    The algorithm takes experienced throughput, buffer condition, and previous downloaded chunk sizes as input, processes them with a convolutional neural network, and produces an optimal bitrate selection in terms of an bitrate-centric objective QoE model.
    It should be noted that Pensieve and FastMPC optimizes the same objective function.
    \item Rate-Distortion Optimized Streaming (RDOS): RDOS is a novel ABR algorithm, which optimizes the weighted combination of an state-of-the-art QoE model KSQI~\cite{duanmu2019ksqi} and negative bitrate to encourage bitrate saving.
    KSQI takes the heterogeneity of source videos, video codecs, and viewing device into consideration by using advanced video quality assessment models~\cite{li2016VMAF,rehman2015ssimplus} as the presentation quality measure.
    We present the results using VMAF~\cite{li2016VMAF} as our presentation quality model as it is open source that facilitates reproducible research.
\end{itemize}
We implement the ABR algorithms in dash.js (version 2.9.2)~\cite{iso2012dash}.
We optimize the free parameters of BB, FastMPC, Pensieve, and RDOS across an independent database of training videos generated from 250 high-quality 4K videos and 3040 network traces.
The training data is generated in a similar fashion to ensure the optimality of ABR algorithms on the test set.
For FastMPC, we compress the javascript code directly instead of performing run-length coding on the lookup table.
In our experiment, we find the simplification introduces minimum overhead and the code size is close to the original implementation~\cite{yin2015control}.
To perform feed-forward prediction in the browser, we convert the actor networks of Pensieve and RDOS to Tensorflow.js~\cite{google2018tfjs} and save the models in the client local storage via IndexedDB~\cite{mozilla2018indexeddb}.

\noindent\textbf{Viewing devices:} The ultimate receivers of streaming videos are human beings, who consume multimedia on a large variety of viewing devices.
In this paper, we consider three mostly used viewing devices according to~\cite{netflix2018device}, including full high definition (FHD) monitor, smartphone, and ultra high definition (UHD) TV.
Note that the presentation quality is a function of viewing device, which we take into account with device adaptive VMAF scores.
An alternative presentation quality measure is the SSIMplus index~\cite{rehman2015ssimplus}, which offers richer and more precise device-adaptive scoring, though its implementation is not publicly available.

\noindent\textbf{Experimental setup:} In order to generate meaningful and representative test videos for subjective experiment, we conduct a set of DASH video streaming experiments, recorded the relevant streaming activities, and reconstructed the streaming sessions using video processing tools.
DASH videos were pre-encoded and hosted on an Apache Web server.
We used Mahimahi~\cite{netravali2014mahimahi} to emulate the network conditions from our corpus of network traces, along with an 80 ms round-trip time, between the client and server.
The client video player is a customized Chromium browser (version 73) supporting H.264 and HEVC playback.
Both the client and server run on the same computer with an Intel i7-6900K 3.2GHz CPU.
Before a streaming session is initialized, the player selects one viewing device from FHD monitor, Phone, and UHDTV, and parses presentation QoE scores from the manifest file accordingly.
After each video streaming session, a log file was generated on the client device, including selected bitrates, duration of initial buffering, and the duration of each stalling event.
We then reconstructed each streaming video with FFmpeg~\cite{ffmpeg}.
Aiming to evaluate the performance at steady status, we force all ABR algorithms to start with the same quality level.
We remove the initial buffering and the first chunk from the streaming videos for presentation.

The simulation with realistic network traces and ABR systems ensures the generated streaming videos come from the real-world distribution.
Furthermore, the end-to-end treatment from server, network to client viewing device enables controlled data analysis, which is not possible with only the streaming videos in the wild.

\noindent\textbf{Summary:} A total of 1,350 streaming videos (5 source videos $\times$ 2 encoders $\times$ 9 network traces $\times$ 5 ABR algorithms $\times$ 3 viewing devices) are generated for presentation.
The mean and standard deviation of the video duration are 30.7 and 1.8 seconds, respectively.

\subsection{Subjective Testing}
\noindent\textbf{Choice of testing methodology:}
Given the large-scale streaming videos and the limited capacity of subjective testing, it is prohibitively difficult to employ the pairwise comparison subjective testing method, which arguably produces more reliable ratings~\cite{bt500subjective,ye2014active}.
There exist two alternatives in the literature including single-stimulus (SS) and single stimulus continuous quality evaluation (SSCQE) methods.
In SS methods, a single streaming video is presented and the assessor provides an index of the entire presentation.
The approach has become the standard subjective testing method in the field of visual communication and has been applied in several streaming video QoE datasets~\cite{ghadiyaram2014study,duanmu2016sqi,duanmu2018quality}.
By contrast, the SSCQE scheme records not only continuous-time QoE scores while participants are viewing test stimulus, but also retrospective scores at the end of the presentation.
The past decades has witnessed an increasing trend towards the usage of single stimulus continuous quality evaluation in streaming video QoE assessment\cite{moorthy2012video,bampis2017qoe,bampis2018towards}, thanks to its capability to provide scene-dependent and time-varying quality evaluation.
We conduct a small-scale pilot study to investigate the feasibility of each method in the QoE assessment of streaming videos, based on which we obtain some interesting feedback.
First, participants report that they frequently encounter difficulties in recalling retrospective scores in the SSCQE experiment due to the limited mental capacity.
The phenomenon is evident by the low repeatability of the SSCQE experiment.
Second, there is time delay between the recorded instantaneous quality and the video content, and such delay varies between subjects and is also a function of slider ``stiffness''.
This is an unresolved issue of the general SSCQE methodology, but is avoided when only a single score is acquired.
On the other hand, the long duration of test videos in SS as opposed to the international recommendation~\cite{bt500subjective} comes with a cost.
We find that participants gradually loss interests in viewing test stimuli, merely paying attention to the first few segments.
To overcome these problems, we propose a variant of SS by introducing an auxiliary task in the experiment.
In particular, each subject is asked to (1) perform a keystroke whenever a rebuffering event occurs and (2) provide the overall QoE score at the end of each presentation.
The auxiliary task not only motivates participants focusing on the experiment materials, but also helps us identify outliers who do not attend to the full test stimuli.
We empirically observe a better inner subject correlation in the proposed experiment.
As a result, we adopt the dual-task SS as the subjective testing methodology in the development of WaterlooSQoE-IV.

\begin{figure*}[t]
    \centering
    \captionsetup{justification=centering}
    \subfloat[Picture quality]{\includegraphics[width=0.33\textwidth]{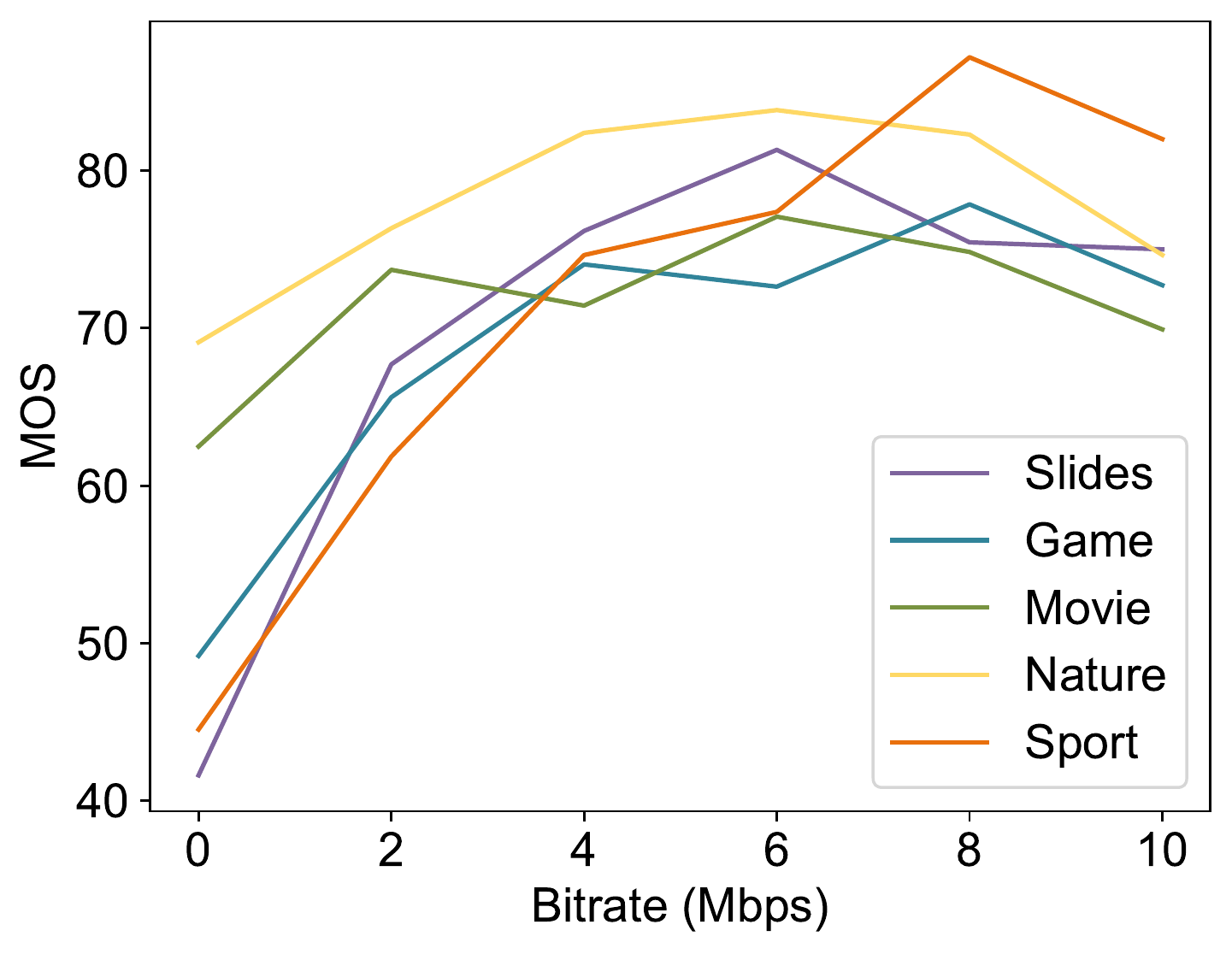}}
    \subfloat[Rebuffering]{\includegraphics[width=0.33\textwidth]{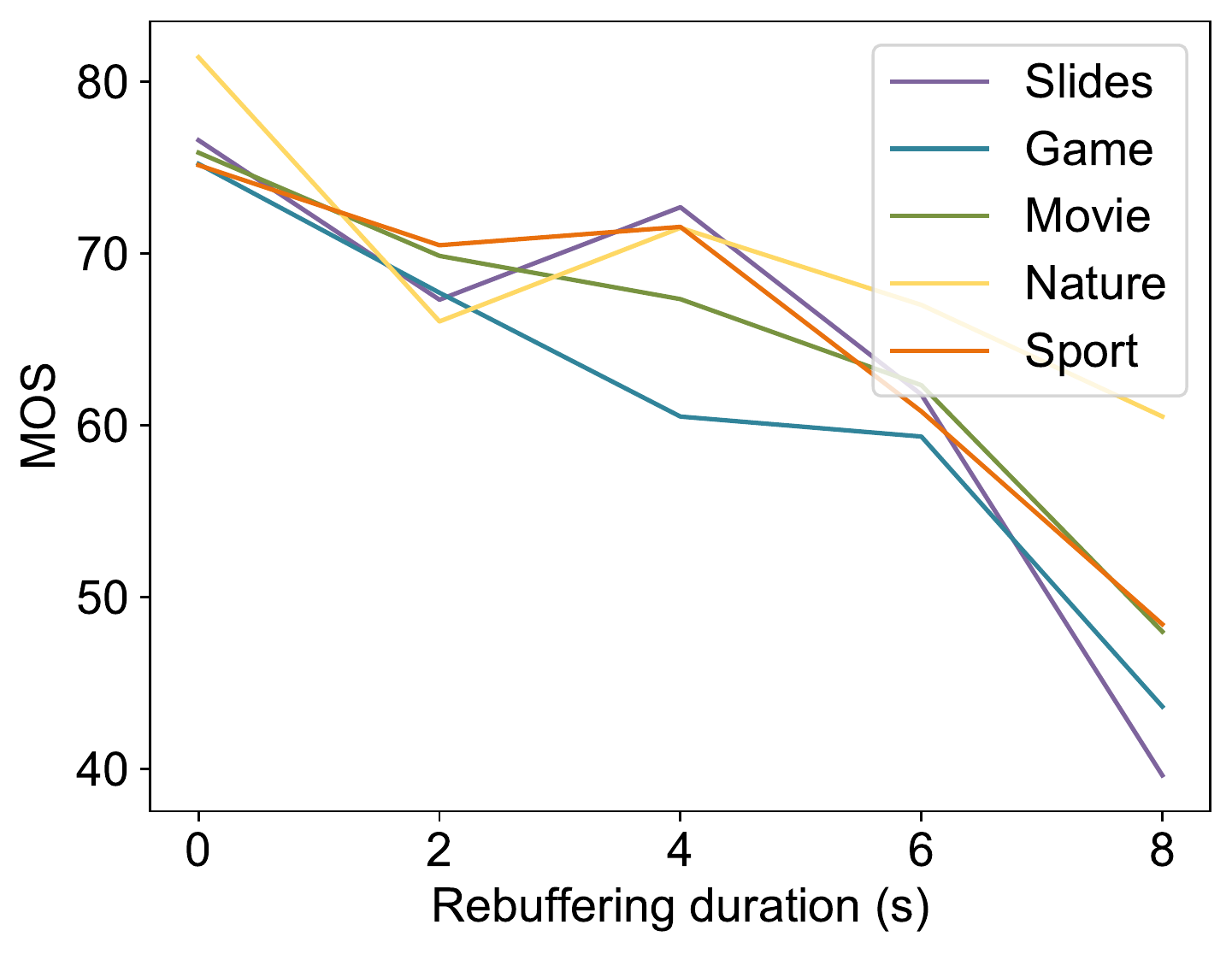}}
    \subfloat[Quality adaptation]{\includegraphics[width=0.33\textwidth]{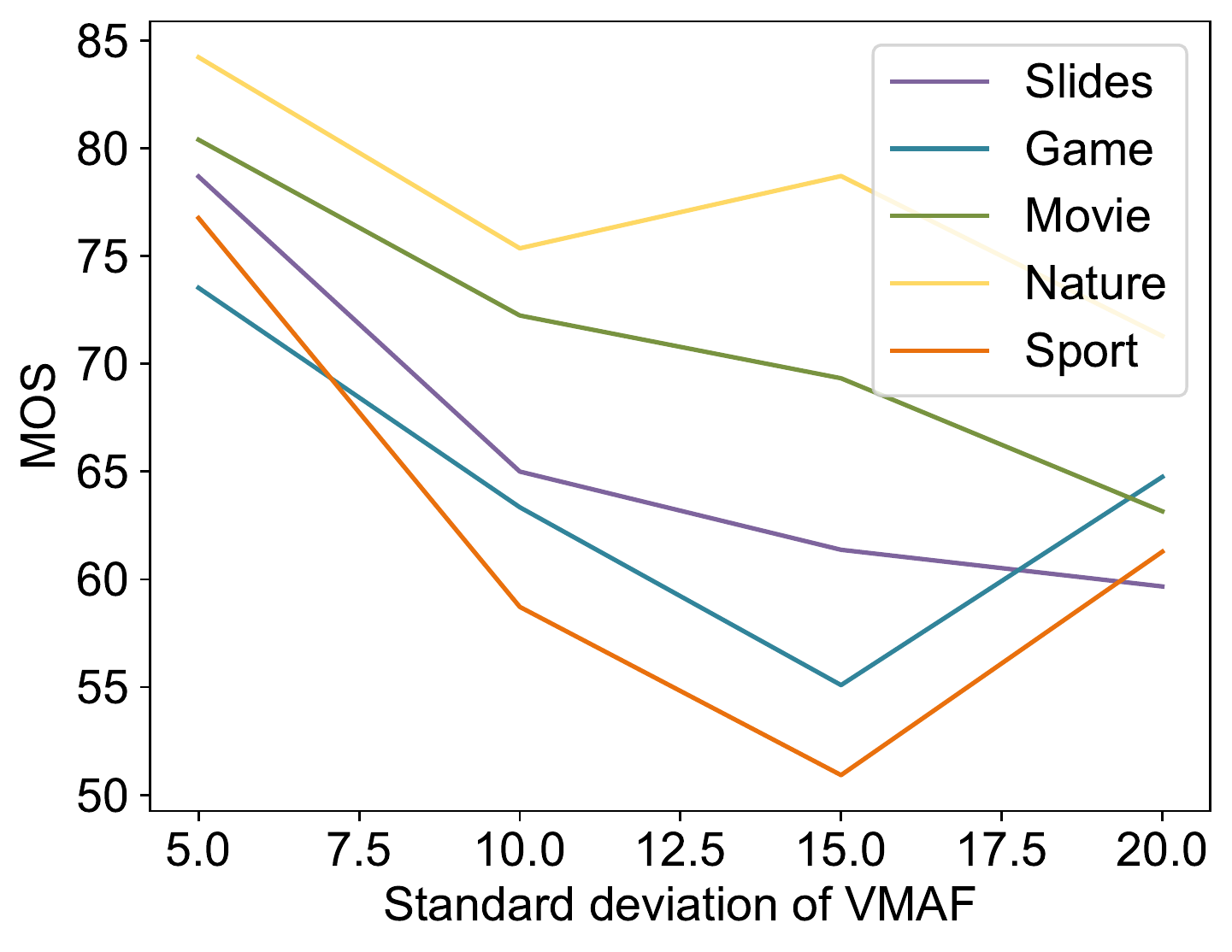}}
    \caption{Influence of source content in quality related factors.}\label{fig:source}
\end{figure*}

\begin{figure*}[t]
    \centering
    \captionsetup{justification=centering}
    \subfloat[Picture quality]{\includegraphics[width=0.33\textwidth]{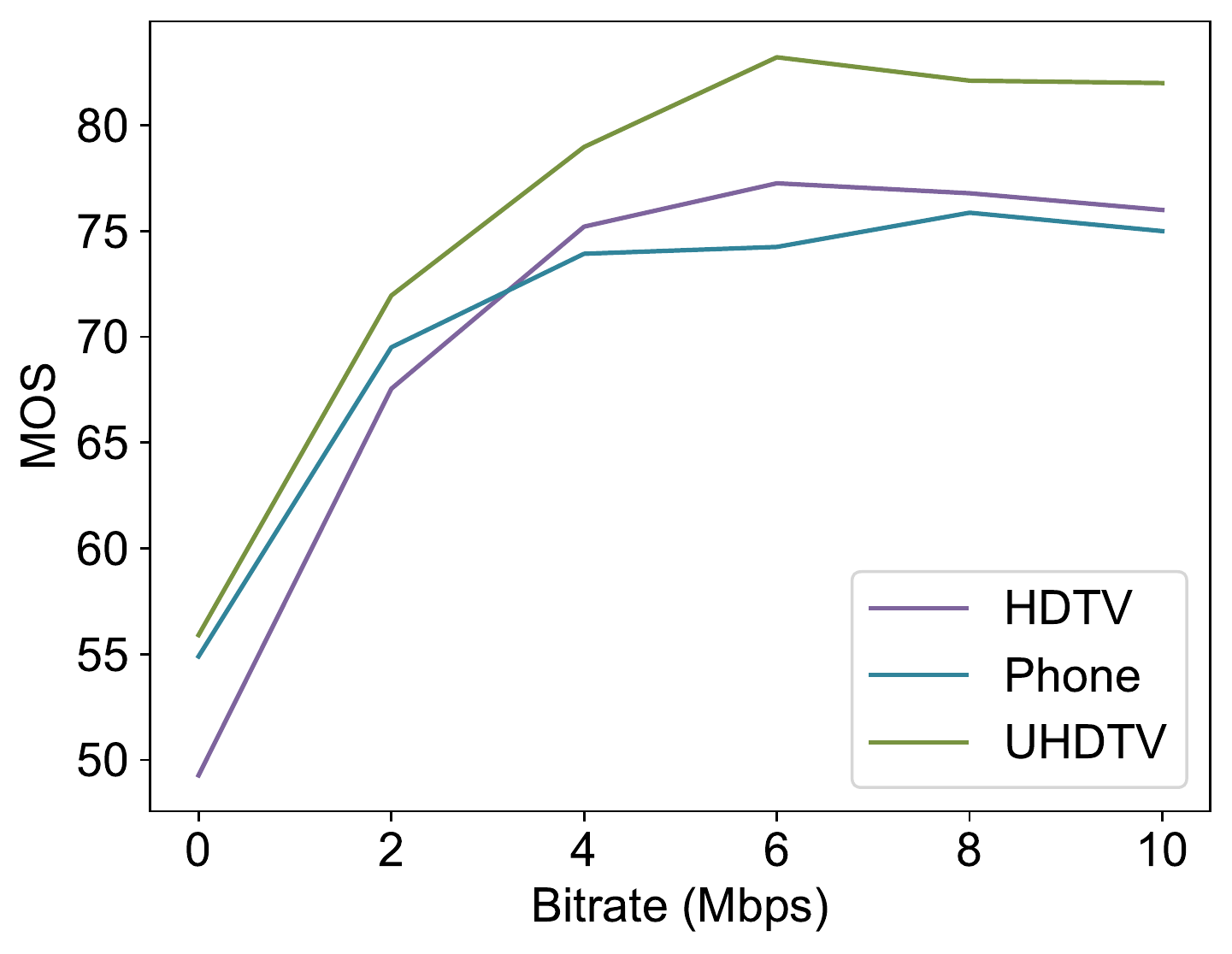}}
    \subfloat[Rebuffering]{\includegraphics[width=0.33\textwidth]{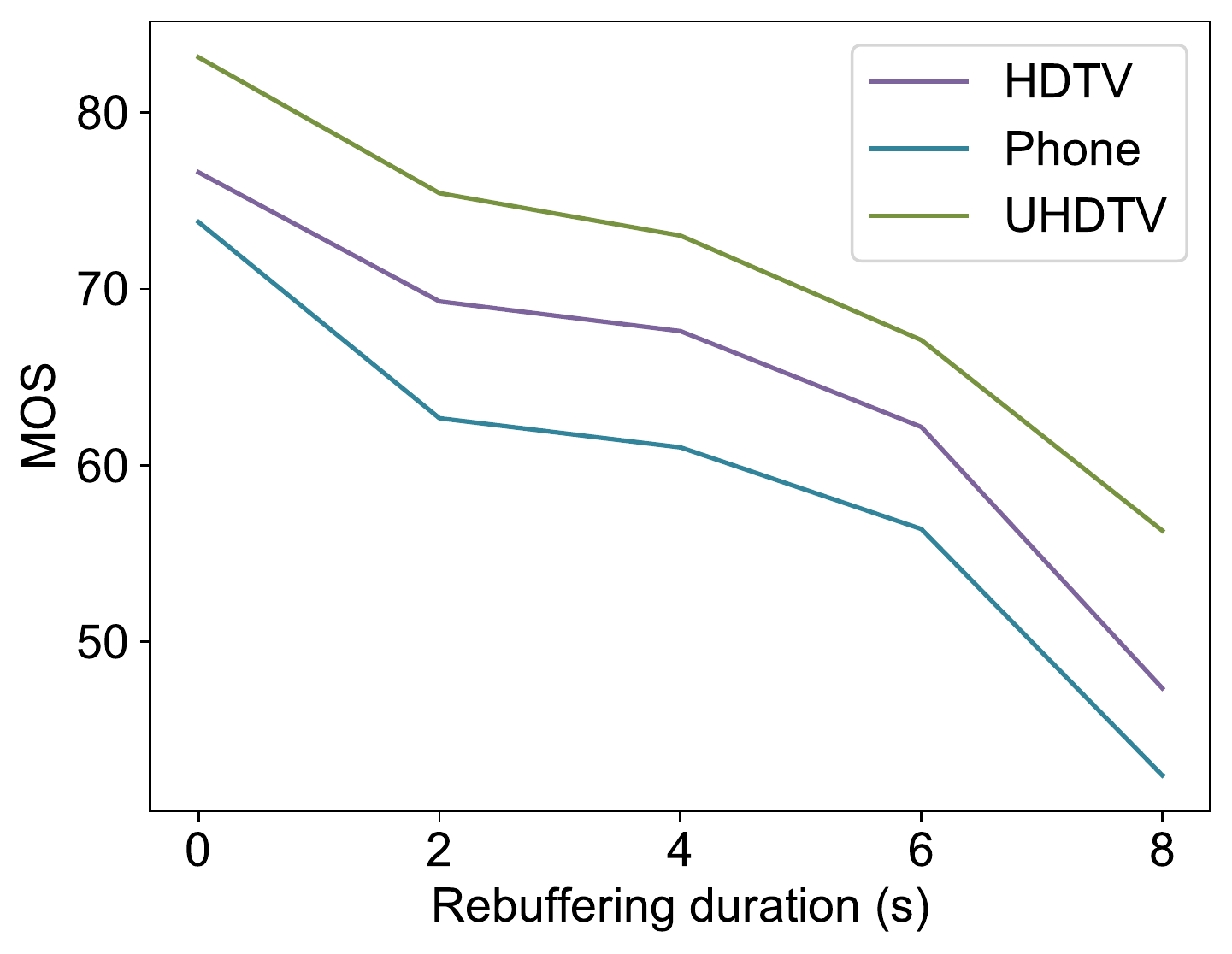}}
    \subfloat[Quality adaptation]{\includegraphics[width=0.33\textwidth]{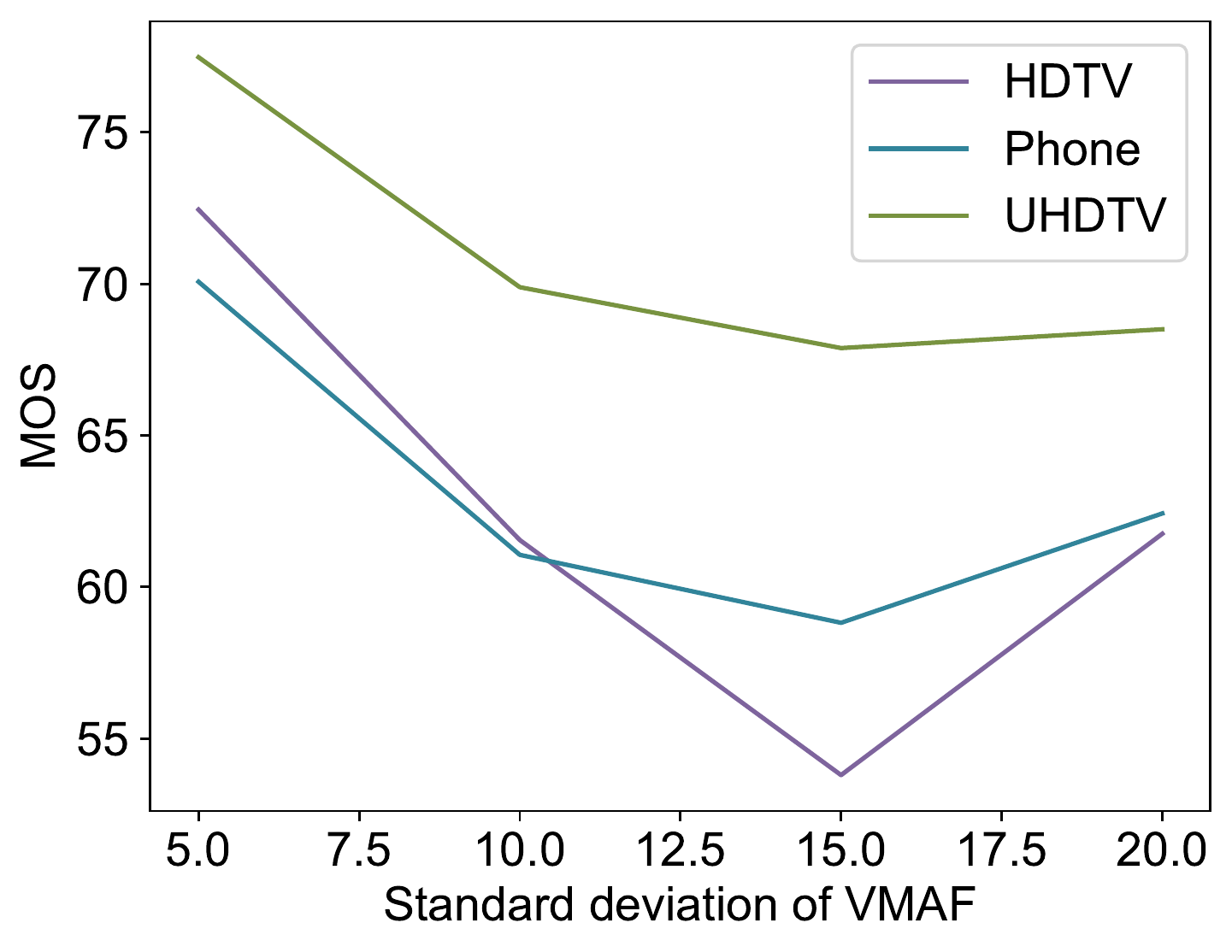}}
    \caption{Influence of device in quality related factors.}\label{fig:device}
\end{figure*}

\noindent\textbf{Experiment procedure:}
The subjective experiment is carried out over a period of eight weeks at the University of Waterloo at Image and Vision Computing subjective testing lab.
The environment is setup as a normal indoor home settings with an ordinary illumination level, with no reflecting ceiling walls and floors.
A customized graphical user interface is used to render the videos on the screen with random order and to record the individual subject ratings on the database.
A total of 97 na{\"i}ve subjects, including 50 males and 47 females aged between 18 and 38, participate in the subjective test.
Given the time constraint, each subject is randomly assigned a viewing device from FHD monitor (24 inch ViewSonic VA2452SM), Phone (5.8 inch Apple iPhone XS Max), and UHDTV (55 inch Sony XBR55X800H).
In the end, the Phone, HDTV, and UHDTV studies received ratings from 33, 32, and 32 participants, respectively.
All videos are displayed at full-screen on each of the devices.
The monitors are calibrated in accordance with the ITU-R BT.500 recommendation~\cite{bt500subjective}.
Observers are seated at a distance of 0.2 m, 1 m, and 4 m from the Phone, HDTV, and UHDTV displays, respectively.
Visual acuity and color vision are confirmed from each subject before the subjective test.
A training session is performed, during which, 8 videos that are different from the videos in the testing set are presented to the subjects.
We used the same methods to generate the videos used in the training and testing sessions.
Therefore, subjects knew what distortion types would be expected before the test session, and thus learning effects are kept minimal in the subjective experiment.
Subjects were instructed with sample videos to judge the overall QoE considering all types of streaming activities in the session.
For each subject, the whole study takes about 5.5 hours, which is divided into eleven sessions spanning over three days.
In order to minimize the influence of fatigue effect, the length of a session was limited to 25 minutes.
The choice of a 100-point continuous scale as opposed to a discrete 5-point ITU-R Absolute Category Scale (ACR) has advantages: expanded range, finer distinctions between ratings, and demonstrated prior efficacy~\cite{sheikh2006statistical}.
Since the eleven sessions were conducted independently, there is a possibility of misalignment of their quality scales.
In order to alleviate the problem, we performed a separate experiment for realignment, where ten videos from each session were collected as test stimuli.
The videos chosen from each session roughly covered the entire quality range for that session.

\noindent\textbf{Post-processing:}
The raw subjective scores are converted to Z-scores.
We remove the ratings of streaming videos where each rebuffering event is not associated with an keystroke.
In addition to the outlier removal scheme suggested in~\cite{bt500subjective}, we remove subjects who failed to accurately perform 10\% of the auxiliary task, leaving 92 valid subjects.
The results of the realignment experiment were used to map the Z-scores to mean opinion score (MOS) in accordance with~\cite{sheikh2006statistical}.
Specifically, we assume a linear mapping between Z-scores and MOS.
The coefficients are learnt by minimizing the prediction residual.
One mapping is learnt for the experiment on each day and applied to the Z-scores of all videos in the respective sessions to produce the realigned MOS for the whole database.
The standard deviation of opinion scores and the mean Spearman's rank-order correlation coefficient (SRCC) between individual subject ratings and the MOSs are 17.35 and 0.67, respectively.

\section{Subjective Data Analysis}
In this section, we analyze the collected subjective data in WaterlooSQoE-IV to reveal the relationships between various factors and subjective QoE.
We present one of the first attempts in exploiting the user heterogeneity of QoE with respect to each quality related factors.
In addition to a quantitative analysis of ABR algorithms, we evaluate the performance of objective QoE models.

\subsection{Interactions between Subjective QoE and Various Factors}
\noindent\textbf{Source content:} Source content has a statistically significant impact in the visual QoE on the WaterlooSQoE-IV database based on the result of an analysis of variance (ANOVA) test.
The average opinion scores of Slides, Game, Movie, Nature, and Sport are 61, 62, 69, 74, and 62, respectively.
To investigate how source content affects perceived picture quality, we draw the subjective ratings of each content with respect to the average bitrate in Figure~\ref{fig:source} (a), wherein we removed the samples with rebuffering duration longer than one second.
We find that the contents with high temporal complexity such as Sport generally exhibit a very low presentation quality at low bitrate region while a relatively high quality can be attained at 7 Mbps bitrate.
On the other hand, the spatially composite videos (\textit{e.g.} Nature) benefit little from excessive bitrate, although a decent quality can be achieved with little resources.
Figure~\ref{fig:source} (b) shows the influence of source content in the perception of rebuffering.
It can be observed that for the stalling at the same temporal instance and of similar duration, human subjects tend to give a higher
penalty to the video with a higher motion complexity and a more coherent story line.
The result confirms the findings of~\cite{liu2015deriving}.
We perform a similar analysis on the relationship between viewing device and perception of quality adaptation.
Given that bitrate cannot account for the heterogeneity in source content, we employ an advanced perceptual video quality assessment model VMAF as the presentation quality measure in the subsequent analysis.
The subjective ratings with respect to the segment-level VMAF variation is given in Figure~\ref{fig:source} (c), from which we can see that humans are more sensitive to the quality variations in the temporally complex contents such as Sport and Game.
This phenomenon is also orally confirmed by the participants.

\noindent\textbf{Encoder type:} The influence of encoder type turns out to be statistically insignificant on the proposed dataset.
The MOSs of H.264 and HEVC video sequences are 66 and 65, respectively.
The result is somewhat surprising since it is well-known that HEVC exhibits a better rate-distortion performance than H.264.
By taking a closer look at the encoded video streams, we find HEVC produces 12\% higher segment-level bitrate variation on average, leading to more rebuffering events for the ABR algorithms who do not have knowledge about the chunk sizes in the future.
The result has significant implications to the video content distributions.
From a content provider standpoint, an encoder with better rate-distortion performance does not always transfer to a higher QoE, especially when the rate control is not taken with caution.
The interaction between each two operation points in the communication pipeline should be considered in the practical content distribution.
From an algorithm developer standpoint, chunk size varies significantly from a Gaussian distribution in HEVC, which has been commonly used in the development of ABR algorithms.
The next generation ABR algorithms may benefit from chunk size look ahead and a better rate model to combat the uncertainty in the chunk size.

\noindent\textbf{Viewing device:} While it has been widely accepted that subjective QoE of streaming videos is a function of viewing condition, their quantitative relationship has yet to be explored.
With the diversity of device types in the WaterlooSQoE-IV, we are able to investigate the influence of viewing device in QoE for the first time.
The average ratings on HDTV, Phone, and UHDTV are 64, 64, and 69, respectively, adhering to the conventional belief in the connection between display devices and QoE.
The effect of viewing device in subjective ratings is shown to be statistically significant in an ANOVA test.
We further explore how viewing device influences the quality related factors.
Figure~\ref{fig:device} (a) illustrates the interaction between bitrate and viewing device without streaming videos longer than one second, from which we have several observations.
First, for each device, the perceptual quality increases monotonically with the bitrate, while the maximum quality varies with respect to the viewing devices.
Second, UHDTV consistently receives the highest reward given the same bitrate resource, which may be a consequence of a longer viewing distance.
It is also worth noting that both phone and HDTV has a bitrate region in which it receives a higher QoE ratings than the other device.

\begin{figure}[t]
\centering
\includegraphics[width=0.8\linewidth]{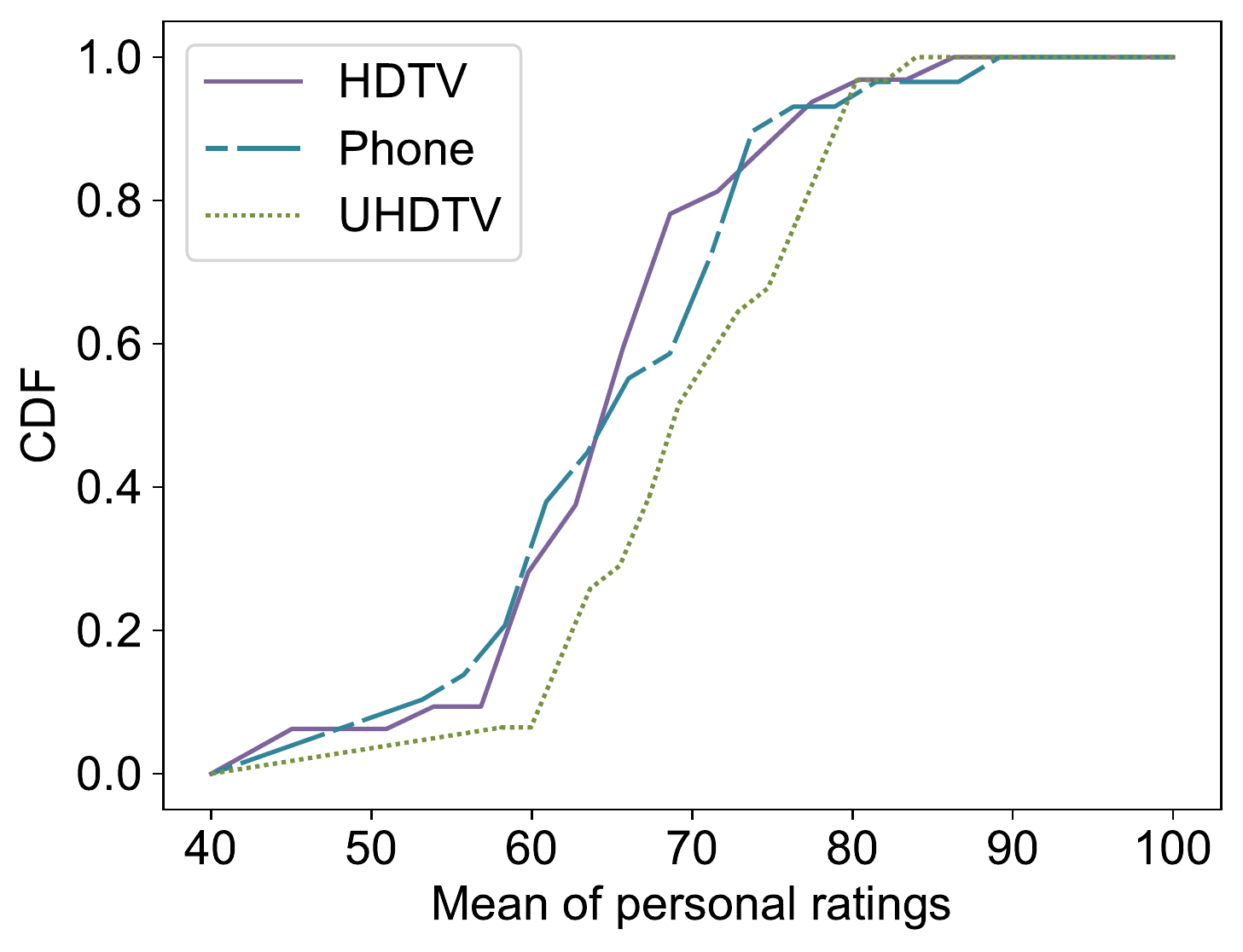}
\caption{Cumulative distribution function of average personal ratings.}
\label{fig:mmos}
\end{figure}

We then carry out an analysis on how viewing device affects the rebuffering experience.
To eliminate the influence of other quality factors, we only consider the streaming videos with an average VMAF around 80 ($\pm$ 10) and remove the ones with significant quality variation (VMAF standard deviation larger than 10).
Figure~\ref{fig:device} (b) shows the relationship between the rebuffering duration and the MOS on three devices.
We summarize the key observations as follows.
First, the subjective QoE decreases monotonically with the rebuffering duration, but the curve starts flattening out at 2 seconds.
This could be explained by duration neglect effects~\cite{hands2001recency}, which assume subjects tend to be insensitive to the duration of a long lasting video impairment.
Participants loss their patient at 6 seconds when the rate of decreasing in MOS starts to increase.
Second, rebuffering events introduce $\sim$10\% and $\sim$18\% more QoE degradation on Phone than on HDTV and UHDTV across all range of rebuffering durations, respectively.
This phenomenon was not observed in previous studies.
One explanation may be that the short viewing distance of smartphone provides a more immersive viewing experience, and thus the interruption caused by rebuffering make subjects more frustrated.

\begin{figure*}[t]
    \centering
    \captionsetup{justification=centering}
    \subfloat[Sensitivity to low quality vs. sensitivity to rebuffering]{\includegraphics[width=0.33\textwidth]{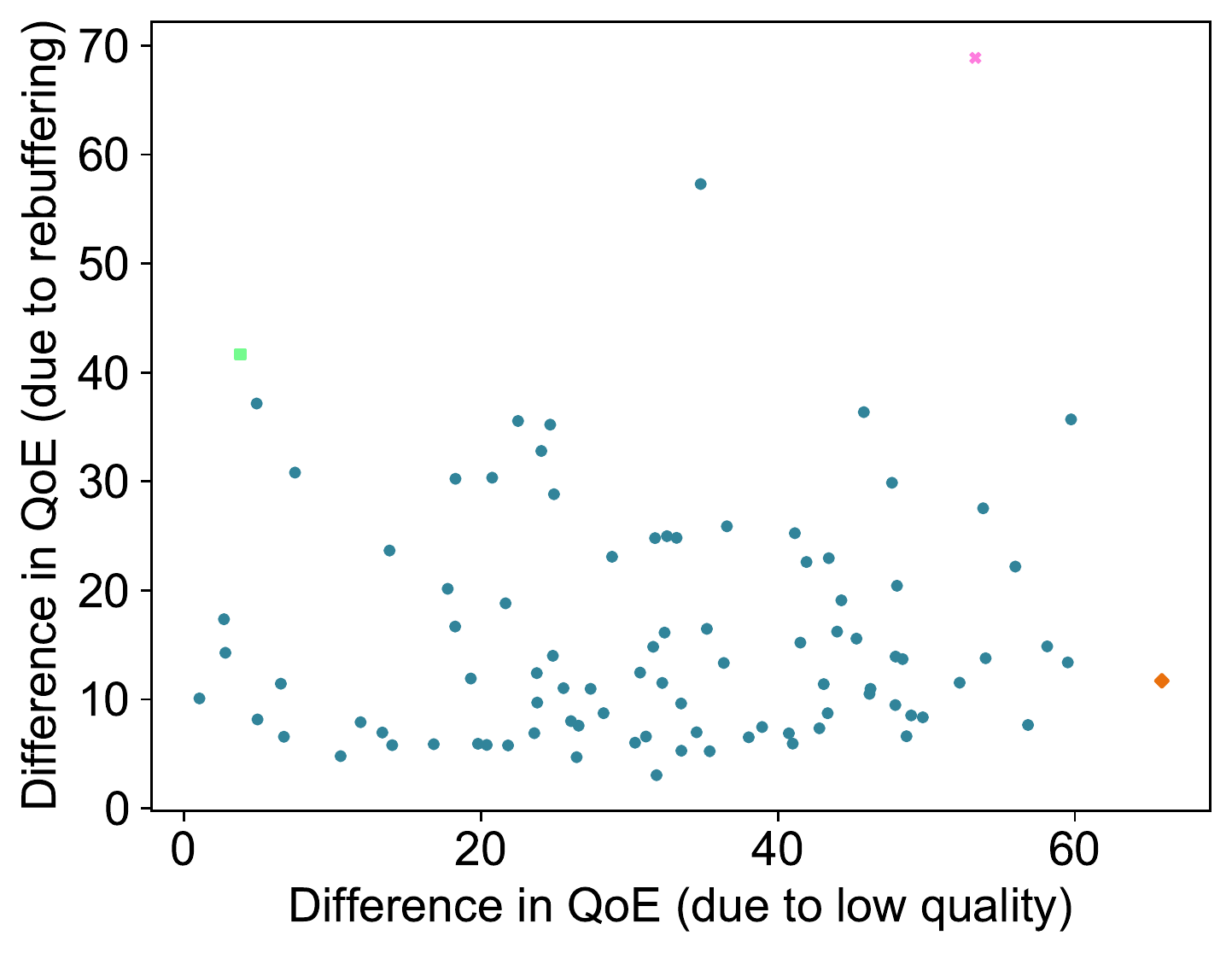}}
    \subfloat[Sensitivity to low quality vs. sensitivity to quality adaptation]{\includegraphics[width=0.33\textwidth]{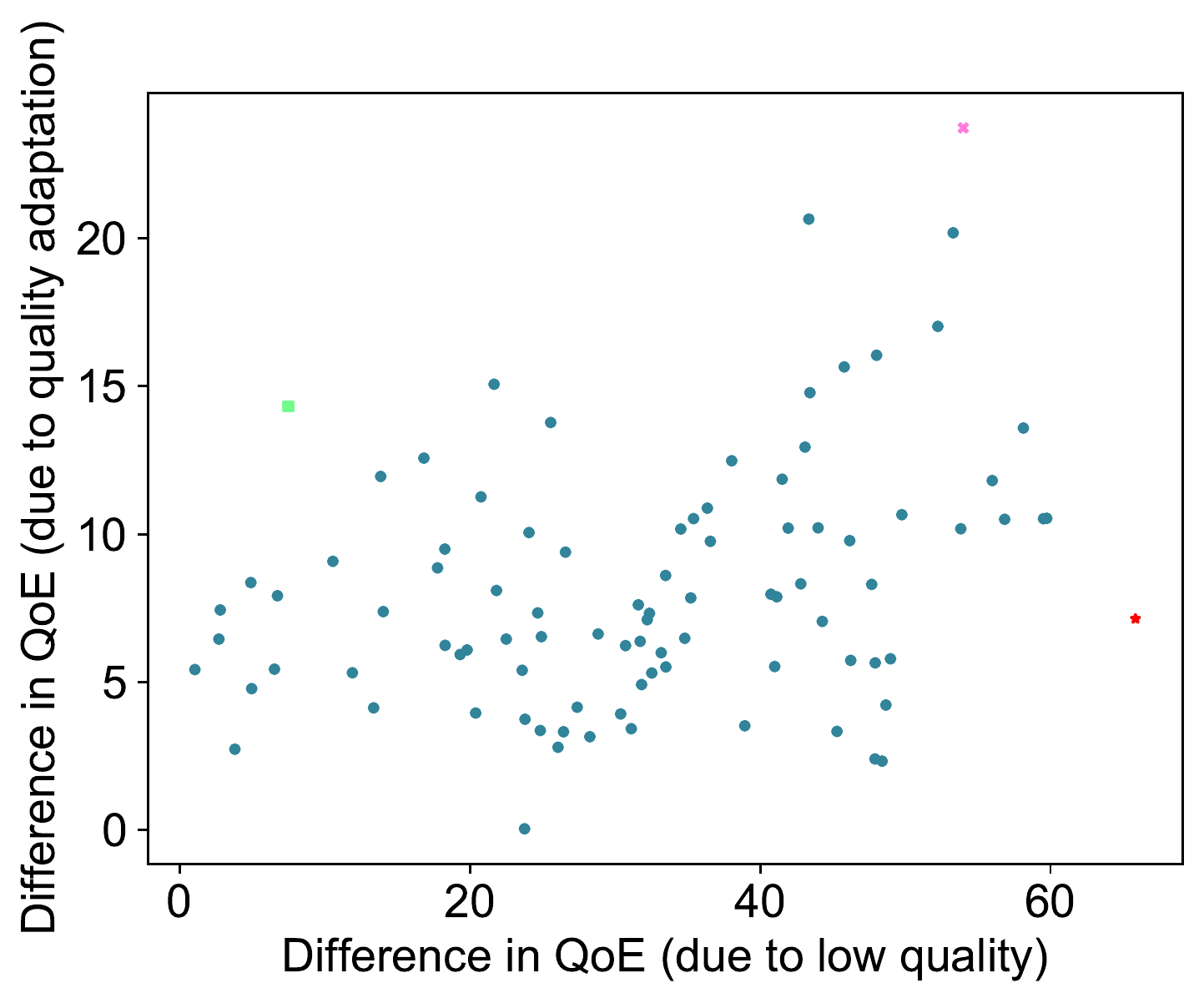}}
    \subfloat[Primacy effect vs. recency effect]{\includegraphics[width=0.33\textwidth]{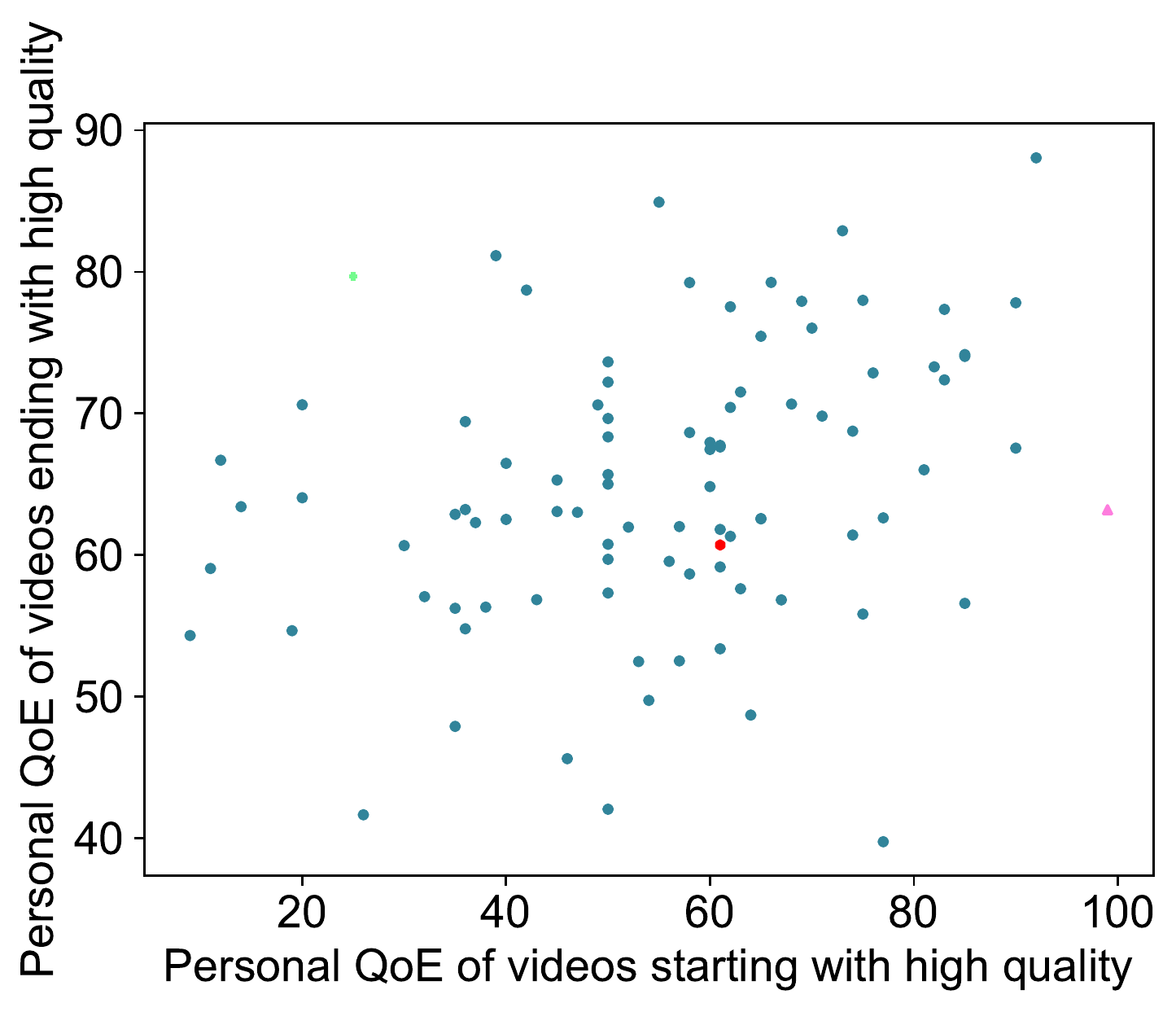}}
    \caption{User heterogeneity with respect to different dimensions. }\label{fig:user}
\end{figure*}

Last, we present a similar analysis towards quality adaptation.
We discard the samples with rebuffering duration greater than one second to get rid of the influence of stalling.
The relationship between the quality variation and MOS on three testing displays is shown in Figure~\ref{fig:device} (c), from which we have several interesting observations.
We first notice that QoE does not decrease monotonically with respect to the magnitude of quality variation.
By examining the streaming activity logs, we find the streaming videos with large quality variation (standard deviation of VMAF $>= 20$) contains $\sim$25\% more positive adaptations than the other quality variation levels, suggesting that positive adaptations are preferred over negative adaptations.
In addition, subjects are more sensitive to small quality variations on HDTV than on Phone, while the trend is reversed at large quality adaptations.
Meanwhile, quality adaptation tends to have very little impacts on UHDTV.
The underlying mechanism remains to be explored in the future. 

\subsection{User Heterogeneity}
Although it has long been assumed that different users exhibit considerable heterogeneity in QoE, there has been limited work studying the user-level QoE in streaming videos~\cite{wang2016data,gao2020personalized,huo2020meta}.
It remains an open question how subjects differ from each other in quality perception of streaming videos.
Given the abundant QoE ratings in the WaterlooSQoE-IV database, we acquire an unique opportunity to present an in-depth analysis to this problem.
We identify two types of differences across individuals as follows.

\noindent\textbf{Bias:} We compute the average rating of each subject as an indicator of his/her overall sensitivity to distortions.
The personal average scores are comparable within each device group because each group of viewers watched identical stimuli under the same experiment setting.
Figure~\ref{fig:mmos} shows the distribution of the average personal QoE ratings across all subjects on each viewing device.
We see that the bottom 10\% of users on average are 20\% more tolerant to the same distortion level than the top 10\% of users.
To verify the difference in user perception is significant, we perform the Wilcoxon signed-rank test on the ratings from the top 10\% of users and the bottom 10\% of users.
User heterogeneity is consistently significant on all three devices.

\begin{figure*}[t]
\centering
\includegraphics[width=0.95\linewidth]{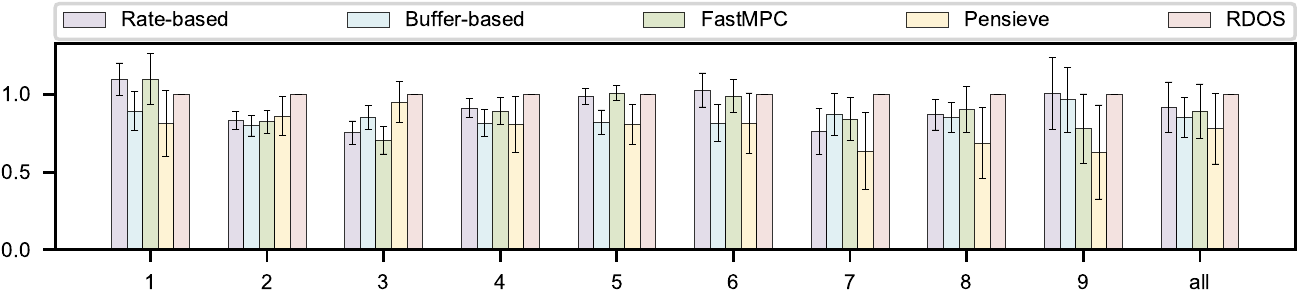}
\caption{Performance of ABR algorithms on each testing network trace. Results are normalized against the performance of RDOS. Error bars span $\pm$ one standard deviation from the average.}
\label{fig:abr_network_plot}
\end{figure*}

\begin{figure*}[t]
\centering
\includegraphics[width=0.95\linewidth]{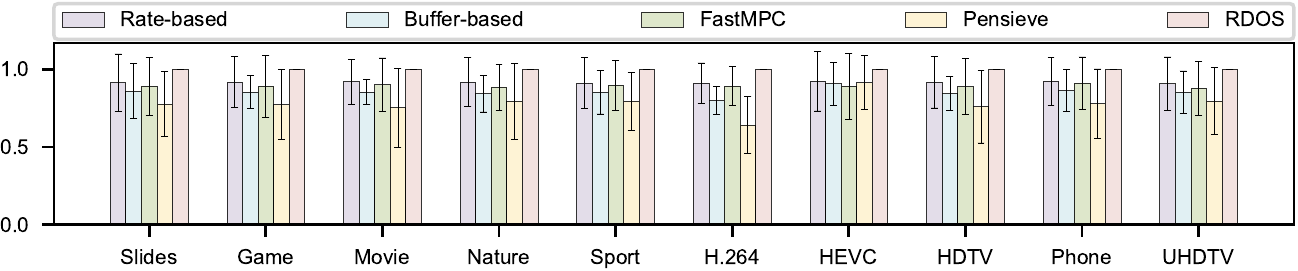}
\caption{Performance of ABR algorithms on each content, video codec, and viewing device. Results are normalized against the performance of RDOS. Error bars span $\pm$ one standard deviation from the average.}
\label{fig:abr_factor_plot}
\end{figure*}

\noindent\textbf{Preference:} We investigate the viewers' sensitivity to each type of distortions.
We first detail the experiment setup in our analysis of the rebuffering experience.
To eliminate the influence of other quality factors, we only consider the streaming videos with an average VMAF around 80 ($\pm$ 10) and remove the ones with significant quality variation (VMAF standard deviation larger than 10).
We then partition the QoE ratings of each user $i$ into two groups including the ratings on videos without rebuffering and the ratings on videos with a rebuffering event longer than one second, where the two sets are denoted as $\mathcal{Q}_{\bar{r}}^i$ and $\mathcal{Q}_r^i$, respectively.
The individual sensitivity to rebuffering is measured as
\begin{equation}
    s_r^i = \frac{1}{|\mathcal{Q}_{\bar{r}}^i|}\sum_{x \in \mathcal{Q}_{\bar{r}}^i} x - \frac{1}{|\mathcal{Q}_{r}^i|}\sum_{y \in \mathcal{Q}_{r}^i} y,
\end{equation}
where $|\cdot|$ represents the cardinality of a set.
To analyze the sensitivity of video quality, we construct two sets of user ratings $\mathcal{Q}_{q}^i$ and $\mathcal{Q}_{\bar{q}}^i$, where $\mathcal{Q}_{q}^i$ contains the ratings on videos with average VMAF greater than $60$ and $\mathcal{Q}_{\bar{q}}^i$ is the complement set of $\mathcal{Q}_{q}^i$.
Streaming videos with rebuffering duration longer than one second and VMAF standard deviation larger than 10 are discarded from both sets to single out the influencing factor.
We compute the individual sensitivity to presentation quality with
\begin{equation}
    s_q^i = \frac{1}{|\mathcal{Q}_{q}^i|}\sum_{x \in \mathcal{Q}_{q}^i} x - \frac{1}{|\mathcal{Q}_{\bar{q}}^i|}\sum_{y \in \mathcal{Q}_{\bar{q}}^i} y.
\end{equation}
The analysis in quality variation is performed in a similar fashion.
We collect two groups of videos, with and without VMAF standard deviation greater than 10.
No rebuffering events appear in both set of samples.
The average VMAF score of the videos in the groups are 80.
We denote corresponding subjective ratings of each cluster as $\mathcal{Q}_{a}^i$ and $\mathcal{Q}_{\bar{a}}^i$, respectively.
The individual sensitivity to quality adaptation is defined as
\begin{equation}
    s_a^i = \frac{1}{|\mathcal{Q}_{a}^i|}\sum_{x \in \mathcal{Q}_{a}^i} x - \frac{1}{|\mathcal{Q}_{\bar{a}}^i|}\sum_{y \in \mathcal{Q}_{\bar{a}}^i} y.
\end{equation}
To ensure that the result is statistically meaningful, each set contains at least $30$ videos.

One of the most important questions in the adaptive streaming is whether to switch to lower quality or stay to the same quality level and stall when there is a network congestion.
Figure~\ref{fig:user} (a) compares the individual sensitivities to low quality and to rebuffering, where each dot represents the preference of a subject.
There are two important takeaways from these results.
First, there exists significant variability among subjects in their preferences in the presentation quality and rebuffering.
For example, the subject denoted by green/orange are 40\%/50\% more sensitive to quality/rebuffering, respectively.
Second, there is generally no trade-off between the two types of preference, as a subject can be sensitive to both metrics simultaneously (such as the subject represented with the magenta cross).

Another critical question in the design of ABR algorithms is whether humans prefer shorter durations of high quality content in the midst of a low quality stream, or if they prefer to view the low quality stream without any fluctuation in quality.
Our results in the individual sensitivities to quality adaptation and to low quality reveals that the answer is subject dependent.
Specifically, about 90\% subjects prefer quality adaptation to remaining at low quality as shown in Figure~\ref{fig:user} (b), whereas the other 10\% present the opposite preference.
The point in green/red represents an typical viewer who prefers low quality/adaptation to the other distortion type.
Similar to the results of low quality and rebuffering, a viewer (\textit{e.g.} the subject represented with the magenta cross) can be significantly sensitive to both quality adaptation and low quality comparing to other participants.

In addition to the preference in distortion types, we also explore the viewers' sensitivity to the primacy effect and the recency effect~\cite{ebbinghaus1913memory}.
The primacy effect and the recency effect suggest that there is a cognitive bias which results in a subject recalling information presented toward the beginning and end of a stimulus, respectively.
We examine the effects in the context of adaptive streaming.
Specifically, we collect two sets of videos, in which either the first segment or the last segment has a VMAF score lower than 70.
Again, we exclude the samples with rebuffering events and significant segment-level quality variation.
The average VMAF of each video sequence in both sets is around 85.
Figure~\ref{fig:user} (c) shows the experiment results, from which we can see that 60\% participants are dominated by the recency effect.
We also find there are three types of viewers who are recency effect-dominated (green diamond), primacy effect-dominated (magenta  triangle), and neutral (red hexagon).

To facilitate the research in personalized QoE optimization, the individual ratings are made publicly available.

\begin{figure}[t]
\centering
\includegraphics[width=0.8\linewidth]{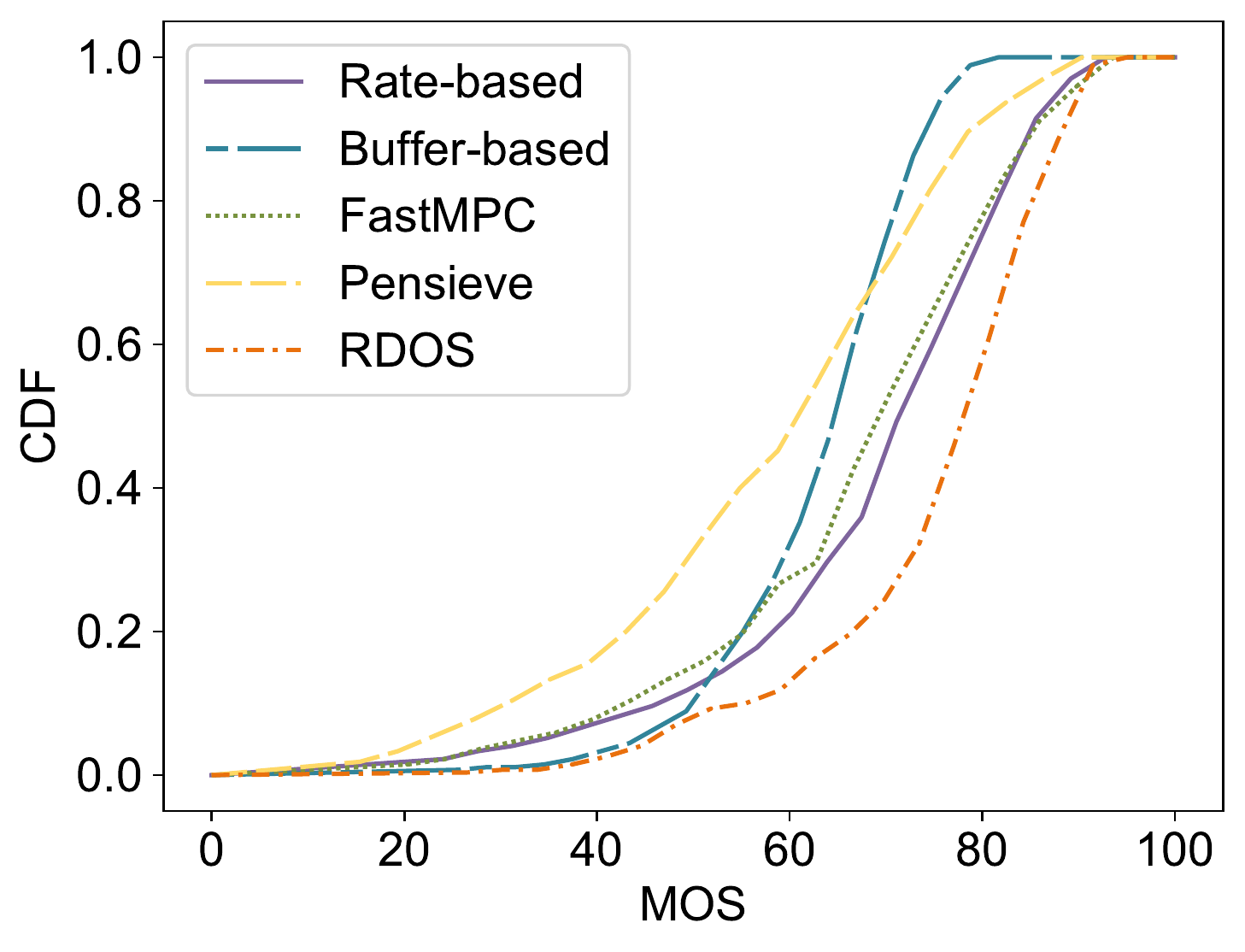}
\caption{Cumulative distribution function of MOS generated from five competing ABR algorithms.}
\label{fig:abr_cdf}
\end{figure}

\begin{table}[t]
\centering
    \caption{Statistical Significance Matrix Based on Wilcoxon-Statistics on the WaterlooSQoE-IV Dataset. A Symbol ``1'' Means That the Performance of the Row Algorithm Is Statistically Better Than That of the Column Algorithm, A Symbol ``0'' Means That the Row Algorithm Is Statistically Worse, and A Symbol ``-'' Means That the Row and Column Algorithms Are Statistically Indistinguishable}\label{tab:wilcoxon_test}
    \scalebox{0.9}{
    \begin{tabular}{c|c c c c c}
    \toprule
      & Rate-based   & Buffer-based & FastMPC & Pensieve & RDOS \\ \hline
      Rate-based     & - & 1 & - & 1 & 0  \\
      Buffer-based   & 0 & - & 0 & 1 & 0  \\
      FastMPC        & - & 1 & - & 1 & 0  \\
      Pensieve       & 0 & 0 & 0 & - & 0  \\
      RDOS           & 1 & 1 & 1 & 1 & -  \\
    \bottomrule
    \end{tabular}
    }
\end{table}

\subsection{Performance of ABR Algorithms}
Figure~\ref{fig:abr_network_plot} and~\ref{fig:abr_factor_plot} show the MOS that each ABR scheme receives across each dimension. Figure~\ref{fig:abr_cdf} provides the cumulative distribution of MOS attained by the ABR algorithms on the WaterlooSQoE-IV database.
There are three key takeaways from these results.
First, we find that RDOS exceeds the performance of the best existing ABR algorithm with a sizable margin on almost all scenarios considered.
RDOS achieves 10\% and 30\% performance gain over the second best algorithm rate-based and its bitrate-driven counterpart Pensieve, respectively.
Second, it is surprising that buffer-based, FastMPC, and Pensieve are inferior to the de facto rate-based algorithm on average.
This is in sharp contrast to the significant gains claimed in existing studies using one or few test samples of hand-picked video clips and network traces, and verified with casual testing.
Our results suggest that a better QoE model, or a better understanding of the human perceptual experiences, is an essential and dominating factor in improving ABR algorithms, as opposed to advanced optimization frameworks, machine learning strategies, or bandwidth predictors, where a majority of ABR research has been focused on in the past decade.
Third, FastMPC outperforms its data-driven counterpart Pensieve, despite the common objective function.
The first interpretation of the phenomenon is that the convolutional neural network architecture cannot well characterize the vast diversity of network conditions, leading to sub-optimal bitrate selection.
Another possible explanation could be that despite a more accurate throughput prediction, the enormous difference between the objective QoE prediction and subjective QoE response results in misplacements of bitrate resources.
The real cause remains unclear because Pensieve only provides implicit throughput prediction.
Fourth, RDOS demonstrates the most notable performance gain at low-bandwidth conditions.
By having a closer look at the streaming logs, we find that RDOS is able to learn a policy that starts with low bitrate level, gradually switches up, and stays at an intermediate bitrate level at poor bandwidth conditions, while other ABR algorithms either constantly makes conservative decisions or erratically switches up and down according to the instantaneous bandwidth estimate or buffer occupancy observations.
The difference may be explained by the perceptually motivated QoE model employed by RDOS, whereby positive adaptations are preferred over negative adaptations.
Fifth, the rate-based algorithm and FastMPC perform at least on par with the best algorithm RDOS on network traces with small variation such as traces 5, 6, and 9, suggesting the (implicit) data-driven throughput prediction does not always lead to the optimal bitrate selection.
In such cases, future ABR algorithms may exploit the connection-level information to reduce the uncertainty of future throughput~\cite{akhtar2018oboe,Yan2020puffer}.
Sixth, although source content and viewing device have relatively little influences, the performance of ABR algorithms varies significantly over different video codecs.
Consequently, the reported gain in the existing studies obtained on a single encoder does not generalize to other settings.
At last, not a single algorithm provides the best perceptual quality under all network profiles.
This suggests that there is still room for future improvement.

\begin{table}[t]
\centering
    \caption{Performance of Objective QoE Models on WaterlooSQoE-IV}\label{tab:obj_qoe}
    \begin{tabular}{c|c c c}
    \toprule
      QoE model & PLCC & SRCC & KRCC\\ \hline
      Mok2011~\cite{mok2012qdash}               & 0.046 & 0.056 & 0.044 \\
      FTW~\cite{tobias2013youtube}              & 0.147 & 0.082 & 0.072 \\
      Xue2014~\cite{xue2014assessing}           & 0.166 & 0.219 & 0.148 \\
      Liu2012~\cite{liu2012case}                & 0.282 & 0.468 & 0.319 \\
      Yin2015~\cite{yin2015control}             & 0.323 & 0.541 & 0.379 \\
      VideoATLAS~\cite{bampis2017atlas}         & 0.675 & 0.670 & 0.480 \\
      P.1203~\cite{itu2017pnats}                & 0.636 & 0.668 & 0.479 \\
      Bentaleb2016~\cite{bentaleb2016sdndash}   & 0.682 & 0.692 & 0.495 \\
      Spiteri2016~\cite{spiteri2016bola}        & 0.685 & 0.662 & 0.461 \\
      SQI~\cite{duanmu2016sqi}                  & 0.717 & 0.690 & 0.504 \\ 
      KSQI~\cite{duanmu2019ksqi}                & \textbf{0.720} & \textbf{0.699} & \textbf{0.575} \\
    \bottomrule
    \end{tabular}
\end{table}

To ascertain the performance difference among ABR algorithms is statistically significant, we carry out a statistical significance analysis.
The evaluation statistic is the Wilcoxon signed-rank test.
The null hypothesis is that the sample produced by a pair of ABR algorithms come from the same distribution.
In particular, it tests whether the distribution of the differences is symmetric about zero (with 95\% confidence).
The results are summarized in Table IX, where a symbol `1' means the row algorithm performs significantly better than the column algorithm, a symbol `0' means the opposite, and a symbol `-' indicates that the row and column schemes are statistically indistinguishable.
It can be observed that buffer-based and Pensieve algorithms are statistically inferior to the na\"ive rate-based algorithm, while RDOS is significantly better than all competing algorithms, confirming the importance of perceptual QoE modeling.

\subsection{Performance of Objective QoE Models}

\begin{table*}[t]
\centering
    \caption{Statistical Significance Matrix based on F-Statistics on the WaterlooSQoE-IV Dataset. A Symbol ``1'' Means That the Performance of the Row Model Is Statistically Better Than That of the Column model, A Symbol ``0'' Means That the Row Model Is Statistically Worse, and A Symbol ``-'' Means That the Row and Column Models Are Statistically Indistinguishable}\label{tab:vartest}
    \begin{tabular}{c|c c c c c c c c c c}
    \toprule
      & FTW & Mok2011 & Liu2012 & Yin2015 & VideoATLAS & Spiteri2016 & P.1203 & Bentaleb2016 & SQI & KSQI \\ \hline
      FTW               & - & - & 0 & 0 & 0 & 0 & 0 & 0 & 0 & 0  \\
      Mok2011           & - & - & 0 & 0 & 0 & 0 & 0 & 0 & 0 & 0  \\
      Liu2012           & 1 & 1 & - & - & 0 & 0 & 0 & 0 & 0 & 0  \\
      Yin2015           & 1 & 1 & - & - & 0 & 0 & 0 & 0 & 0 & 0  \\
      P.1203            & 1 & 1 & 1 & 1 & - & 0 & 0 & 0 & 0 & 0  \\
      VideoATLAS        & 1 & 1 & 1 & 1 & 1 & - & - & 0 & 0 & 0  \\
      Bentaleb2016      & 1 & 1 & 1 & 1 & 1 & - & - & - & 0 & 0  \\
      Spiteri2016       & 1 & 1 & 1 & 1 & 1 & 1 & - & - & 0 & 0  \\
      SQI               & 1 & 1 & 1 & 1 & 1 & 1 & 1 & 1 & - & 0  \\
      KSQI              & 1 & 1 & 1 & 1 & 1 & 1 & 1 & 1 & 1 & -  \\
    \bottomrule
    \end{tabular}
\end{table*}

Using the WaterlooSQoE-IV database, we evaluate the performance of $11$ objective QoE models for adaptive streaming videos.
The competing algorithms are chosen to cover a diversity of design philosophies, including $9$ classic knowledge-driven QoE models: FTW~\cite{tobias2013youtube}, Mok2011~\cite{mok2012qdash}, Liu2012~\cite{liu2012case}, Xue2014~\cite{xue2014assessing}, Yin2015~\cite{yin2015control}, Spiteri2016~\cite{spiteri2016bola}, Bentaleb2016~\cite{bentaleb2016sdndash}, SQI~\cite{duanmu2016sqi}, and KSQI~\cite{duanmu2019ksqi}, and $2$ state-of-the-art learning-based QoE models: VideoATLAS~\cite{bampis2017atlas} and P.1203~\cite{itu2017pnats}.
The implementation for VideoATLAS is obtained from the original authors and we implement the other ten QoE models.
We use mode 0 of P.1203 to ensure that the objective QoE models are evaluated under a comparable setting, where no competing model has the access to the finer (frame) level information.
We have made the implementation of the models publicly available at \url{https://github.com/zduanmu/ksqi}.
For the purpose of fairness, the parameters of all models are optimized on the WaterlooSQoE-I~\cite{duanmu2016sqi} and the WaterlooSQoE-II~\cite{duanmu2017qoe} datasets, except for P.1203~\cite{itu2017pnats} whose training methodology is not specified in the original paper.
For the models with hyper-parameters, we randomly split the datasets into $80$\% training and $20$\% validation sets, and the hyper-parameters with the lowest validation loss are chosen.

Three criteria are employed for performance evaluation by comparing MOSs and objective QoE scores according to the recommendation by the video quality experts group~\cite{vqeg2000metrics}.
We adopt Pearson linear correlation coefficient (PLCC) to evaluate the prediction accuracy, SRCC and Kendell rank correlation coefficient (KRCC) to assess prediction monotonicity.
A better objective QoE model should have higher PLCC, SRCC, and KRCC.

Table~\ref{tab:obj_qoe} shows the PLCC, SRCC, and KRCC on the WaterlooSQoE-IV dataset, where the best performer is highlighted with bold face.
There are four key takeaways from these results.
First, the objective QoE models which employ advanced VQA models as the presentation quality measure generally performs favorably against the conventional bitrate-based QoE models.
In particular, Bentaleb2016 significantly outperforms Yin2015, where the only difference between them is the presentation quality measure.
The reason may be that the quality of streaming videos is highly content- and viewing condition-dependent and bitrate is insufficient in capturing the perceptual distortions introduced by compression.
On the other hand, state-of-the-art VQA models consistently provide meaningful and consistent QoE predictions across video contents, video resolutions, and viewing conditions/devices, making them an ideal choice for presentation quality measure in adaptive streaming.
Second, although the learning-based QoE models perform competitively on their original test sets, they do not perform well on the proposed database.
Specifically, VideoATLAS and P.1203 are even inferior to the linear QoE model Bentaleb2016, suggesting that learning-based models exhibit low generalizability, likely due to the mismatch between the limited training data and the diverse streaming environments.
Third, it is important to account for the interactions among video presentation quality, rebuffering experience, and quality adaptation experience, evident by the notable improvement from Bentaleb2016 to SQI and its variant KSQI.
At last, KSQI delivers the best performance on the dataset. The experimental results demonstrate that a promising direction for further improvement resides in a deeper understanding on HVS and a better way to integrate the prior knowledge with human annotated data.

We carry out a statistical significance analysis by following the approach suggested in~\cite{sheikh2006statistical}.
First, a nonlinear regression function is applied to map the objective quality scores to predict the subjective scores.
We observe the prediction residuals all have zero-mean, and thus the model with lower variance is generally regarded better than the one with higher variance.
We conduct a hypothesis testing using F-statistics.
Since the number of samples exceeds 50, the Gaussian assumption of the residuals approximately hold based on the central limit theorem~\cite{bishop2006pattern}.
The test statistic is the ratio of variances.
The null hypothesis is that the prediction residuals from one quality model come from the same distribution and are statistically indistinguishable (with 95\% confidence) from the residuals from another model.
After comparing every possible pairs of objective models, the results are summarized in Table~\ref{tab:vartest}, where a symbol `1' means the row model performs significantly better than the column model, a symbol `0' means the opposite, and a symbol `-' indicates that the row and column models are statistically indistinguishable.
The result of the statistical significance test confirms the improvement of recently developed QoE models upon the traditional bitrate-centric models, which unfortunately still remains as the major QoE indicator in the development of ABR algorithms.
Specifically, the performance of VideoATLAS, Spiteri2016, P.1203, Bentaleb2016 is noticeably superior to Liu2012 and Yin2015, but generally worse than the best performing algorithms.
The performance of KSQI is the best in the group, winning all competitions with other models.

\section{Conclusion and Future Work}
We introduce so far the most comprehensive QoE assessment database containing 1350 subjective-rated streaming videos that are derived from diverse source contents, video codecs, network conditions, ABR algorithms, and viewing devices.
We present an in-depth analysis on the influencing factors of QoE across the communication pipeline.
Based on the experiment results, we assess an extensive variety of of ABR algorithms and objective QoE models.
The novel database allows us to make a series of interesting findings and observations in terms of subjective experiment setup, interactions between user experience and source content, viewing device and encoder type, heterogeneities of user bias and preference, effectiveness and behaviors of ABR algorithms, and performance of objective QoE models.
The dataset is made publicly available to facilitate future research in ABR and streaming QoE.

Future research may be carried out in many directions.
First, although we have made our best effort to construct the largest dataset in the community, the experiment is by no means exhaustive.
It is highly desirable to reduce the complexity of subjective experiment such that an even larger scale experiment can be conducted in a given capacity.
Second, given the promising results of perceptually motivated ABR algorithms and the significant room for improvements of objective QoE models, we believe a better understanding of human visual system and psychological behaviors will further advance the visual communication system.
Third, new machine learning based approaches may be developed using the database, aiming for QoE models with stronger generalization capability.

\bibliographystyle{IEEEtran}
\bibliography{ref} 

\end{document}